\documentclass[letterpaper,twocolumn,10pt]{article}
\usepackage{usenix2019_v3}

\usepackage{graphicx}
\usepackage{amsmath}
\usepackage{msc}
\usepackage{subfig}
\usepackage{booktabs}
\usepackage{multirow}

\setmsckeyword{}
\title{Understanding Realistic Attacks on Airborne Collision Avoidance Systems}

\usepackage{authblk}

\author[1]{Matthew Smith}
\author[1,2]{Martin Strohmeier}
\author[2]{Vincent Lenders}
\author[1]{Ivan Martinovic}

\affil[1]{Department of Computer Science, University of Oxford, UK}
\affil[ ]{\textit{first.last@cs.ox.ac.uk}}
\affil[2]{Cyber Defence Campus, armasuisse Science + Technology, Switzerland}
\affil[ ]{\textit{first.last@armasuisse.ch}}

\begin{document}

\maketitle

\begin{abstract}

Airborne collision avoidance systems provide an onboard safety net should normal air traffic control procedures fail to keep aircraft separated. These systems are widely deployed and have been constantly refined over the past three decades, usually in response to near misses or mid-air collisions. Recent years have seen security research increasingly focus on aviation, identifying that key wireless links---some of which are used in collision avoidance---are vulnerable to attack. In this paper, we go one step further to understand whether an attacker can remotely trigger false collision avoidance alarms. Primarily considering the next-generation Airborne Collision Avoidance System X (ACAS X), we adopt a modelling approach to extract attacker constraints from technical standards before simulating collision avoidance attacks against standardized ACAS X code. We find that in 44\% of cases, an attacker can successfully trigger a collision avoidance alert which on average results in a 590\,ft altitude deviation; when the aircraft is at lower altitudes, this success rate rises considerably to 79\%. Furthermore, we show how our simulation approach can be used to help defend against attacks by identifying where attackers are most likely to be successful.

\end{abstract}
\section{Introduction}

Over the past decade, security research in aviation has highlighted a raft of issues, predominantly around unauthenticated wireless communication channels. These channels enable aircraft-to-ground or aircraft-to-aircraft communications, which in turn supports air traffic control (ATC) surveillance activities. The global aviation infrastructure is moving towards even higher usage of existing and planned wireless links, led by regional and international airspace modernization efforts and specified by the International Civil Aviation Organization's Global Air Navigation Plan \cite{international2007global}.

As its most vital function, ATC keeps aircraft safe by ensuring they are sufficiently separated both laterally and vertically. To fulfil this task, controllers require an accurate surveillance picture of their controlled airspace, which is obtained using several independent voice and data link communications. Despite this redundancy and highly professionalized services, there are still occasions where aircraft end up too close to each other and need to be deconflicted immediately. In such situations, safety systems onboard the aircraft called \textit{collision avoidance systems} (CAS) use avionic data links to automatically communicate with the nearby aircraft. This process then leads to CAS issuing compulsory instructions to the pilots of each aircraft in order to deconflict the situation.

The decision to use existing ATC communication signals for CAS purposes was led by the US Federal Aviation Administration (FAA) in 1981 and developed into the Traffic Alert and Collision Avoidance System (TCAS). Historically, the need for the introduction and improvement of airborne CAS has evolved from near misses or mid-air collisions of aircraft. Consequently, both technology and rules have seen changes, for example in response to the Überlingen disaster of 2002 \cite{GFBAAI2004}, after which following CAS instructions was made compulsory for pilots. Some aircraft even automatically follow deconflicting measures, flying the necessary manuevers directly and without pilot input~\cite{Botargues2009,Eurocontrol2017a}.
           
With the move towards both digitalization and automation in the air, the involved systems are increasingly subject to scrutiny by the security community. This is illustrated by a growing body of literature examining the underlying wireless communication links (a recent survey is found in \cite{strohmeier2020securing}).

Since all current and next generation CAS rely on unauthenticated wireless links such as Automatic Dependent Surveillance -- Broadcast (ADS-B) or the Secondary Surveillance Radar (SSR) Modes A, C and S, the input for CAS is widely considered insecure. However, a comprehensive security analysis explicitly examining airborne collision avoidance has been notably absent thus far. To fill this crucial gap in the literature, we conduct a comprehensive analysis of the real-world impact of wireless attacks on CAS, specifically the next-generation Airborne Collision Avoidance System X (ACAS X), which is due to be rolled out globally in this decade~\cite{SesarJU2020}. 

In our analysis, we focus on the requirements and feasibility of wireless attacks on airborne CAS. By investigating both theoretical and practical environmental conditions, we can illustrate which types of attack, and which outcomes, are possible under realistic constraints. Based on this framework, we develop a simulator-based approach to test attacks on official code from the ACAS X standard. The input to our simulator is provided by 6000 randomly sampled real-world trajectories extracted from the OpenSky Network, which were collected from six airports in several countries~\cite{Schafer2014}.

Our work demonstrates that an adversary can successfully conduct attacks on state-of-the-art airborne CAS but that system design ensures that the ability to have fine-grained control over a target aircraft is restricted in the real world. The practical investigations show that an attacker can trigger a CAS alert in 44\% of our simulated runs rising to 79\% when we focus on aircraft flying at vulnerable lower altitudes. On average, attacks caused a 590\,ft deviation from the original, non-interfered trajectory. Finally, our analysis provides novel insights into effective detection and mitigation of such attacks, e.g. the identification of vulnerable areas around airports.

We make the following contributions:
\begin{enumerate}
    \itemsep-2pt
\item We provide theoretical bounds on successful CAS attacks and the requirements placed on a capable attacker.
\item We construct a novel collision avoidance threat simulator based around officially standardized ACAS X code, fed with real-world aircraft trajectories from the OpenSky Network. Using the simulator, we identify the impact of realistic attacks on targeted aircraft.
\item Based on modelling and simulator analysis, we consider the most effective countermeasures to CAS attacks, including concrete recommended procedures to identify and mitigate the most vulnerable parts of the airspace.
\end{enumerate}

In the next section, we cover the relevant background and related work. We outline our threat model in Sec.~\ref{sec:threat} before exploring the theoretical constraints on attackers in Sec.~\ref{sec:bounds}. We describe our simulator approach in Sec.~\ref{sec:simulate} and explore the results from this in Sec.~\ref{sec:results}. Finally, we offer key insights in Sec.~\ref{sec:discuss}, countermeasures in  Sec.~\ref{sec:countermeasures}, and conclude in Sec.~\ref{sec:conclude}. Our test data and ACAS X simulation environment source code will be made available on publication.

\section{Background \& Related Work}
\label{sec:background}

Although CAS exist in a range of domains, the unique composition of aviation infrastructure, particularly the existence of a central controller in ATC, directly impacts the way these systems work and how they are used. In this section, we cover the relevant background, highlighting important systems and how they fit together. We also note that this paper focuses on technologies used in non-general aviation (GA). GA aircraft are heterogenous with regards to equipment, whereas we are interested in aircraft which are required to use collision avoidance. Specifically, since 2005, ICAO have mandated that all aircraft with a take off mass exceeding 5700\,kg or permitted to carry more than 19 passengers must be equipped with a collision avoidance system~\cite{ICAO2006a}. 

\subsection{Air Traffic Control Surveillance}

ATC is tasked with continuously keeping the aircraft safe. Typically this is achieved by having human Air Traffic Control Operators (ATCOs) manage \textit{controlled airspace} by directly issuing instructions to aircraft. A range of classifications exist, with \textbf{A} being the highest class, where only aircraft flying by instrument are permitted~\cite{ICAO2018a}.

ATCOs monitor aircraft separation using surveillance technologies. These comprise a range of tools, but most fundamentally Primary Surveillance Radar (PSR) and Secondary Surveillance Radar (SSR). PSR is a `traditional' radar, usually comprising a spinning receiver and transmitter, which sends a high-power beam and monitors round trip time and angle of arrival~\cite{Saul-Pooley2017psr}. Whilst effective, it provides a limited amount of information to an ATCO---just the location of aircraft. To supplement this, ATCOs use SSR, a cooperative surveillance mechanism whereby ATC interrogate transponders on board aircraft. These transponders reply with some information determined by the transponder mode.

\paragraph{Mode A/C}
The oldest mode still in operation, Mode A responds with a locally assigned transponder code called a \textit{squawk} and Mode C with the aircraft's pressure altitude when interrogated~\cite{ICAO2004ssr_modes}. Usually these modes are used together, providing an ATCO with both radar position, altitude and identity information. As airspace has become busier, Mode A/C alone is no longer sufficiently to effectively manage aircraft.

\paragraph{Mode S}
A more recent development is Mode S, which uses a similar interrogation mechanism but is a more versatile data link over which ATC can request much more information. Here, the ground radar interrogates an aircraft for a specific piece of information, such as its current altitude, heading or airspeed as recorded by the aircraft, as well as `selected' altitude pilots have set for the flight management system~\cite{ICAO2008, ICAO2004ssr_modes}. Furthermore, this mode uses a global identifier known as a \textit{Mode S} or \textit{ICAO} address. This is hard coded into the transponder and is assigned to a specific aircraft. This mode allows ATCOs to not only see what an aircraft is doing, but also its intention.

\paragraph{Automatic Dependent Surveillance---Broadcast}
In the past decade, aircraft have begun to adopt the latest surveillance tool known as Automatic Dependent Surveillance---Broadcast (ADS-B). One of the main features is the independent transmission of surveillance data by an aircraft over the Mode S channel~\cite{strohmeier2014a}. This allows ATC to monitor aircraft without constantly interrogating them, especially by exploiting onboard GPS to report position. However, it is not intended to fully replace Mode S, rather help reduce channel congestion~\cite{Schafer2014}. Importantly, one way ADS-B can be implemented is through an extension of the Mode S transponder called Mode S Extended Squitter. This allows longer packet sizes over Mode S to enable ADS-B reporting.

\subsection{Collision Avoidance Systems}
Despite ATC's best efforts, sometimes aircraft become too close and are on a collision course. CAS provide an automated safety net for aircraft and have been in use since the 1990s. In this section, we cover the two main CAS implementations and relevant information on how CAS is used by pilots.

\subsubsection{CAS Concepts}
In aviation, CAS leverage existing surveillance technologies---primarily Mode S---to establish nearby aircraft and detect potential collisions. Despite ACAS X being the next generation of TCAS, its core functionality and aims are similar. 

A simple collision avoidance scenario contains two aircraft: the \textit{ownship} and the \textit{intruder}. In such a scenario we take the point of view of the ownship, with the intruder being some nearby aircraft potentially on a collision course. To measure proximity to the intruder, the ownship has two main phases of surveillance when using Mode S:
\begin{enumerate}
    \itemsep-0.75pt
    \item Passive surveillance: the default, where CAS listens for air-ground surveillance link responses from nearby aircraft and estimates their proximity.
    \item Active surveillance: when an intruder has an altitude within 10000\,ft of ownship and is horizontally within around 3\,NM or 60 seconds of the ownship.
    At this point, ownship will directly interrogate the intruder in a similar way to a Mode S ground station~\cite{ICAO2006d}. As part of this phase, the two aircraft may communicate their intended CAS actions, i.e. to climb or descend, to reduce the chance of both choosing a similar action.
\end{enumerate} 

\begin{figure}[]
    \centering
    \includegraphics[width=0.9\columnwidth]{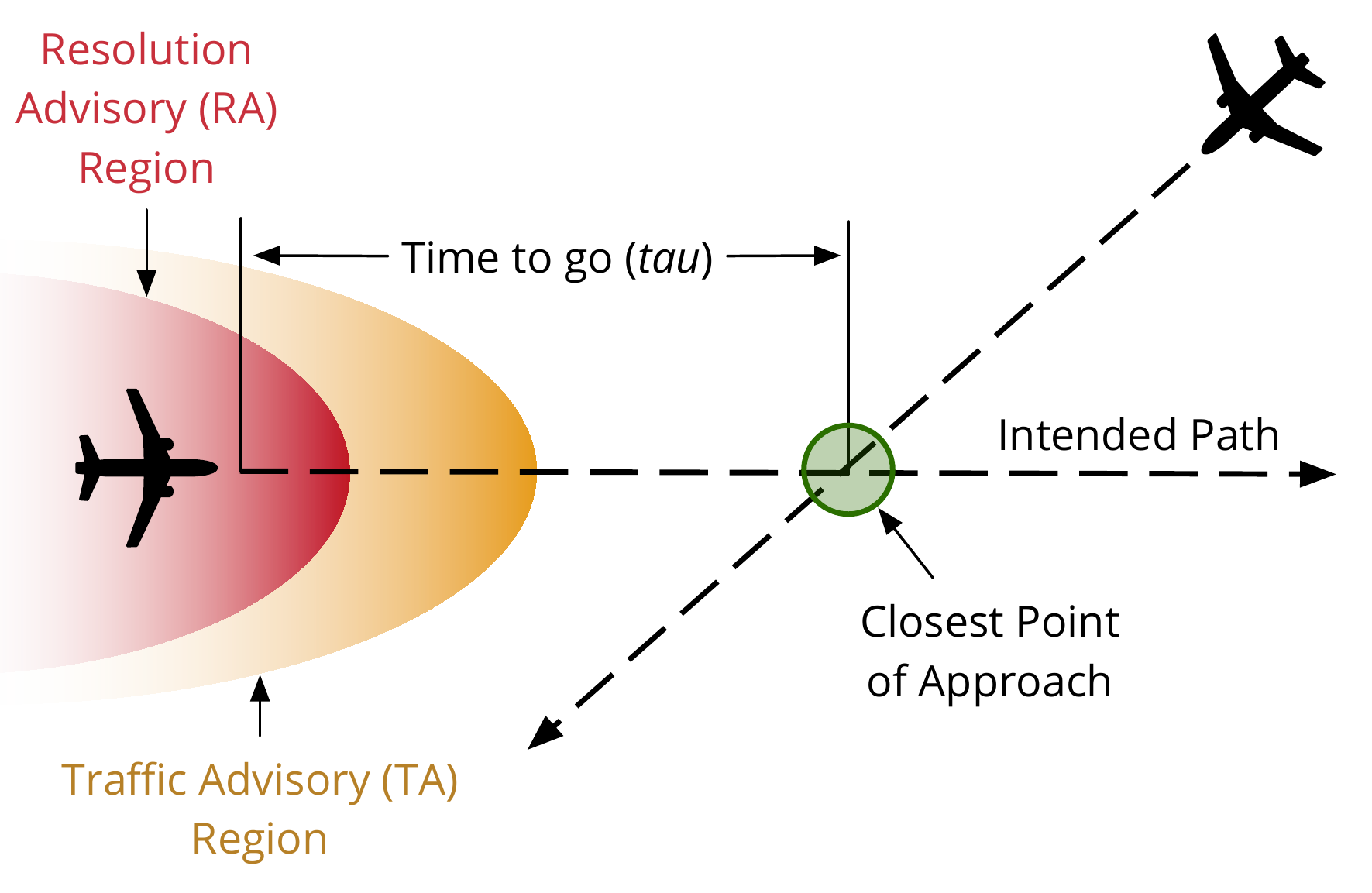}
    \caption{Key collision avoidance system concepts including the protected volume, \textit{tau} and the closest point of approach.}
    \label{fig:tcas-explainer}
\end{figure}

Crucially, the proximity to nearby aircraft is represented not by distance but by \textit{tau}, the time-to-go until the closest point of approach (CPA) to the ownship~\cite{FAA2011b}. If the CPA falls within the volume protected by CAS, then the intruder is considered a threat. We provided a representation of a protected volume in Fig.~\ref{fig:tcas-explainer} and cover the concept of \textit{tau} in more detail in Sec.~\ref{sec:bounds}.

Depending on the speed and altitude of the intruder and ownship, CAS sets time-to-go boundaries at which it triggers alerts. These fall into two groups:
\begin{itemize}
    \itemsep-0.75pt
    \item Traffic Advisory (TA): an indication that an intruder is a potential threat.
    \item Resolution Advisory (RA): an instruction aiding flight crew to increase separation from threats. This will take the form of climbing, leveling off, or descending~\cite{ICAO2006c}.
\end{itemize}
Both TAs and RAs are aurally announced in cockpit, with the RA announcement including specific instructions on the required action~\cite{FAA2011g}. Some aircraft will also provide visual guidance on the required RA maneuver.

\subsubsection{TCAS}

The current CAS implementation in use is the Traffic Alert and Collision Avoidance system, or TCAS. The system has existed in some form since the early 1980s, with usage being mandated worldwide in the aftermath of the 1997 Charkhi Dadri mid-air collision~\cite{Lahoti1997}. The most up-to-date version of TCAS is TCAS II version 7.1, which provides more complex logic surrounding RAs, including the ability to reverse them in the middle of a collision avoidance encounter.\footnote{We refer to this version as TCAS for the rest of this paper.} This was in part a result of the \"{U}berlingen crash in 2002~\cite{GFBAAI2004a, FAA2011c}. 

TCAS requires a Mode S transponder, but can also monitor nearby aircraft using Mode A/C~\cite{FAA2011d}. It can issue RAs with intruder aircraft only equipped with Mode A/C, but these are not coordinated. Using surveillance input, it uses an extensive set of rules to monitor nearby aircraft and the threat they pose, using altitude-based sensitivity levels and a series of thresholds for time-to-go both vertically and horizontally~\cite{FAA2011e}. 

Over time, the system has been modified to support more complex encounters: \textit{multi-threat encounters}, where RAs are coordinated between more than two aircraft, \textit{sense reversals}, where an RA direction is reversed mid-encounter, and \textit{weakening RAs} to help reduce displacement~\cite{FAA2011f}.

\subsubsection{ACAS X}
The next generation of collision avoidance is the Airborne Collision Avoidance System X (ACAS X), which is in the process of being finalized as standard DO-385~\cite{RTCA2018}. Eventually, this system will replace TCAS and although it uses TCAS performance as a baseline, it provides a number of advantages over the older system~\cite{Eurocontrol2017, Castle2012}. 

One of the key improvements is to replace TCAS's rule based logic for a cost function, implemented using optimized threat logic lookup tables~\cite{Castle2012b}. TCAS was seen to have become too complex to easily adapt, so the new approach instead has its logic defined by some rules, which are used with dynamic programming to generate extensive cost tables~\cite{Eurocontrol2017}. Furthermore, ACAS X provides standardized cost tables and a Julia implementation, as opposed to the pseudocode of TCAS~\cite{RTCA2008, RTCA2018}.

Another key change is the move towards long-term adaptability. The system is designed with a `plug and play' approach towards surveillance mechanisms, allowing it to better support ADS-B or future links~\cite{Castle2012c}. It also has a range of modes supporting more complex encounters and purely passive ADS-B based CAS~\cite{Eurocontrol2017}. 

\subsubsection{Collision Avoidance in the Cockpit}

Alerts provided by a CAS are very serious and must be treated with high priority. Flight crew are required to respond to RAs quickly and accurately; ICAO expect that pilots respond within five seconds of the first RA and 2.5 seconds of any subsequent RAs~\cite{ICAO2006b}. Such a requirement exists primarily as a result of the \"{U}berlingen accident in 2002; DHL Flight 611 and Bashkirian Airlines Flight 2937 aircraft collided mid-air when one followed a TCAS RA to descend whilst the other ignored a climb RA and instead descended as instructed by ATC. Had both followed their respective TCAS RAs, they would have separated~\cite{GFBAAI2004a}. Prior to this incident, the precedence of alerts from collision avoidance systems was unclear with respect to ATC instructions. One of the recommendations from the investigation was to make pilot response to RAs compulsory unless doing so would endanger the aircraft, as well as never responding in the opposite direction to an issued RA~\cite{GFBAAI2004}. 

Situations in which CAS appears to be behaving abnormally are less well-defined. In a simulator study looking at how pilots react to attacks on TCAS, results indicated that if the flight crew believe the system to be malfunctioning they are likely to reduce the sensitivity to only issue TAs, or even switch the system to standby~\cite{SmithNDSS2020}.

\subsubsection{Collision Avoidance for Air Traffic Controllers}
Although not the focus of our work, it is important to understand the role that ATC play in collision avoidance encounters. When an aircraft is responding to an RA, ATC must let the aircraft carry out the instruction issued by CAS until the pilots report that they are clear of conflict~\cite{ICAO2016}. As such, RAs can have knock-on impact for the wider airspace particularly in busy regions. A collision avoidance encounter between two aircraft could require ATC to adjust the clearances for many other aircraft not involved. Finally, ATC can be made aware of CAS encounters automatically. TCAS transmits RA information over the 1030\,MHz link, which can be received on the ground and contains information on the active RA~\cite{FAA2011g}.

\subsection{Related Work}

As pointed out by a recent review article on the security of various wireless communication links in aviation \cite{strohmeier2020securing}, there is very little research yet concerning the security of collision avoidance systems despite long-standing suspicion of its derived insecurity (discussed, for example, in \cite{Strohmeier2017}). As one explanation for this lack of scrutiny, the authors explain further that, thankfully, no real-world security incidents related to the deployed TCAS have been (publicly) reported. it is safe to assume that the appearance of such reported incidents would be strongly motivating and accelerating such research. 

Very recently, some researchers from the academic and the hacking community have started to give more thought to the problem of current TCAS security. Berges has looked at the issue from a practical point of view, analyzing the required technical capabilities required for an attack on collision avoidance systems on live aircraft \cite{berges2019exploring}. The work argues that SDR-equipped attackers can successfully trick a target aircraft to track an attacker-generated aircraft and provides a GNU-Radio based implementation of such a threat against open source ADS-B/Mode S decoders. Based on attack-tree analysis, this attack may lead to near mid-air collisions but neither the strong physical layer requirements nor the behavior of the TCAS standard implemented in modern aircraft are considered.

Regarding CAS attacks, Munro has described the fundamental methods used for TCAS spoofing, discussing also the mitigating factors such as pilot performance and suggesting further impact tests in flight simulators \cite{Munro2020}. Such an impact assessment has recently been performed in a large-scale study with 30 pilots \cite{SmithNDSS2020}. Looking at collision avoidance as well as other safety-critical systems, the study found that real-world attacks can successfully create significant control impact and disruption through missed approaches, avoidance maneuvers and diversions. Similarly, they further increased workload, distrust in the affected system.

Our work bridges the gap between fundamental wireless attack primitives on the unsecured radio standards and the impact of possible attack scenarios on collision avoidance systems in general. We present a case with regards to the requirements of attacks on the next generation of collision avoidance systems, ACAS X, based on theoretical and practical analysis of this novel standard and its implementation.

\section{Threat Model}
\label{sec:threat}
The consequences of attacking aircraft systems can be severe and unpredictable, especially since airspace is a complex system. As such, we consider attackers who are determined to cause disruption or erode safety of aircraft. Since the attacks presented in this paper are quite complex, they require specialist equipment and knowledge of the systems.

Typically, an attacker would be located in a static position within 50\,km of an airport in order to be able to successfully carry out the attack. For reasons discussed in Sec.~\ref{sec:bounds}, CAS attacks do not allow an attacker to `control' an aircraft into arbitrary climbs and descends. Instead we presume that an attacker is trying to induce dangerous situations moving aircraft towards busy airspaces, other aircraft or stormy weather. They may aim to force pilots to turn the system off on account of spurious alarms. They could target either one or many aircraft but would need to be in transmission range of any targets.

It is important to note that an attacker will not have full control over the wider system when carrying out a CAS attack. Even if they attacked multiple aircraft simultaneously, these aircraft will also communicate with each other using CAS. The outcome of this is hard to predict, meaning that an attacker can only guarantee a starting role in the effects of their attack. 

\vspace{-7pt}\paragraph{Equipment} For hardware, an attacker would need a commodity SDR such as the Hack RF, which would be coupled with an amplifier and antenna capable of transmitting at 500\,W on 1030/1090\,MHz, and a method to move the antenna to track the aircraft~\cite{RTCA2011}. Such antenna and amplifiers are unlikely to be available off the shelf as they would operate on restricted frequency bands. These items would cost more as a result. Custom software would be needed to listen for and respond to interrogations, manage message sequencing and monitor target position. This would need to be coupled with SDR software to encode and transmit Mode S messages.

\section{Bounding CAS Injection}
\label{sec:bounds}
 
While flooding may deny service, injection allows for a subtle attack that is hard to detect. We now identify the bounds for an attacker to successfully inject a realistic ghost aircraft, assuming our attacker wants to inject a false aircraft which has a sufficiently realistic trajectory to cause a CAS alarm. 

\subsection{Collision Avoidance Basics}

To estimate whether the ownship is on a collision course with another aircraft, both TCAS and ACAS X use \textit{tau}, a calculation of the number of seconds until the CPA between the aircraft based on current observations. \textit{Range tau} is the horizontal time until CPA,  calculated using Eq.~\ref{eq-tau}~\cite{RTCA2018a,RTCA2008a}.

\begin{equation}
    \text{\textit{tau}} = -\frac{r-\frac{\text{SMOD}^{2}}{r}}{\text{min}(-6,rdot)}
    \label{eq-tau}
\end{equation}
  
Here, \textit{r} is the slant range between the interrogating and interrogated aircraft in nautical miles (NM), \textit{SMOD} is a horizontal distance modifier defined as 3\,NM, and \textit{rdot} is the closure speed in 
knots (kt). 

The way in which \textit{tau} is used differs between TCAS and ACAS X. For TCAS, \textit{tau} is compared against thresholds according to a sensitivity level from one to seven. Flight crew usually set the sensitivity level to `automatic' which then allows the current level to be defined by altitude; at each level, there is a specific \textit{tau} threshold for TAs and RAs. If an intruder aircraft produces a \textit{tau} below a TA or RA threshold, an alert will be issued. ACAS X uses a more complex method, taking \textit{tau} values as an input into a cost table lookup. The result is used to determine if an alarm should be sounded and how ownship should respond. This makes the outcome of a given \textit{tau} value harder to predict and thus harder to identify which ownship positions are more vulnerable. However, by using the common methods of TCAS and ACAS X we can calculate some bounds which apply to both systems~\cite{Eurocontrol2017}.

\vspace{-7pt}
\subsection{Attacking Collision Avoidance Systems}

One of the main challenges in attacking CAS is overcoming range and bearing calculations. Both are determined by the ownship based on response characteristics, so it is difficult for an attacker to inject an aircraft with an arbitrary range or bearing. However, for Mode S, altitude is reported by the interrogated aircraft so the attacker can at least arbitrarily control this. For the attack have a predictable effect, the attacker must respond to a sequence of ownship interrogations appearing to be an aircraft on a collision course. This will force the CAS logic on board the target to believe there is a potential intruder nearby. 

To trigger an alarm, the attacker must make the ownship believe that there is some aircraft flying towards it; this requires the attacker to transmit a series of interrogation responses during which the distance between the target and the injected aircraft decreases. CAS calculates range based on interrogation round-trip time (RTT). As such, attackers have can either respond to interrogations when they receive it (i.e. as if they are an aircraft following the normal protocol) or they can preemptively respond to interrogations in an effort to control the RTT as perceived by the target aircraft. The ultimate effect of each is the same, namely that the attacker injects false aircraft messages, but there is a trade off between the two. While preemptive response gives a high degree of control over the position of the false aircraft, it is less reliable as it needs precise timing to predict when the interrogation will be and when a transmission is needed to respond; in contrast, responding when the interrogation arrives at the attacker is very reliable but loses control over the RTT and thus the slant distance between the false and target aircraft.

\subsection{Modelling Realistic Response Constraints}

\begin{figure}[]
    \centering
    \includegraphics[width=0.8\columnwidth]{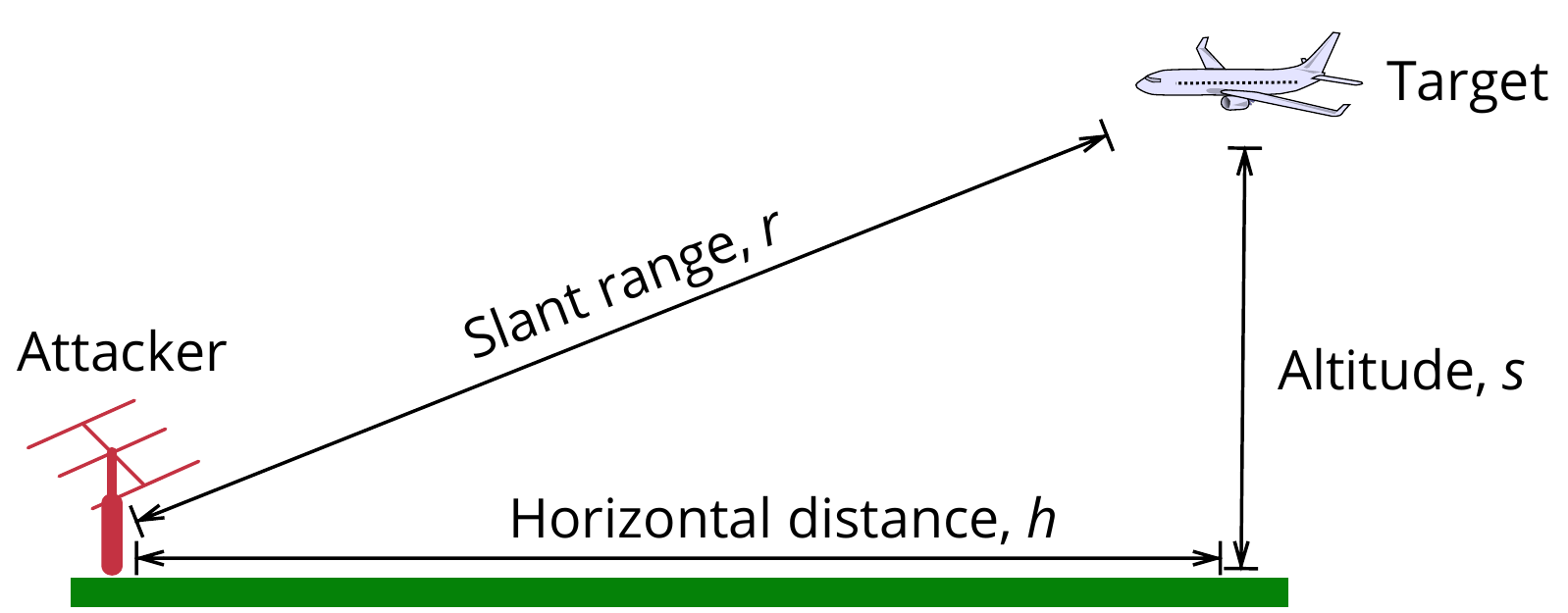}
    \caption{Illustration of the attacker-target aircraft scenario.}\vspace{-10pt}
    \label{fig:att-triangle}
\end{figure}

Our attacker responds to the target aircraft interrogations as we receive them, so they are effectively constrained. To understand how vulnerable aircraft are to  CAS attacks, we identify these constraints and how they affect attack locations. 

First, we model the basic scenario of an attacker injecting a message. Fig.~\ref{fig:att-triangle} illustrates this, with an attacker aiming to inject messages to a target aircraft at slant range $r$. The RTT of an interrogation sets the instantaneous slant range, using $r = \frac{\text{RTT}}{2c}$ where $c$ is the speed of light in a vacuum, and if we consider processing delay to be known and thus removed.

At a given point in time, this forms a right-angled triangle, giving us horizontal distance $h$ and altitude $s$. For an attack to be possible and to give the attacker the best possible chance of success, two conditions on slant distance must be met:
\begin{enumerate}
    \itemsep-1pt
    \item $r$ must be decreasing, and,
    \item $r$ must be shorter than the active surveillance range.
\end{enumerate}
The first condition is met simply by the attacker positioning themselves facing the aircraft path---since arrival and departure patterns around airports are well-defined, this is achievable. The second condition is more complex but can be bounded. Both TCAS and ACAS X define a target to be under active surveillance when \textit{tau} is less than 60 seconds, with ACAS X specifically requiring that the aircraft is airborne or taking off, as well as meeting vertical and horizontal \textit{tau} conditions~\cite{RTCA2018b, RTCA2008a}. For vertical \textit{tau}, it defines
\begin{equation}
    \frac{-(s-4500\,\text{ft})}{min(-1\,\text{ft}/\text{sec}, \dot{s})} \le 60\,\text{sec}
\end{equation}
where $s$ is the altitude difference between the aircraft and intruder and $\dot{s}$ is the rate of change of altitude between the two, which is negative when the altitude difference is decreasing. For horizontal \textit{tau}, it defines 
\begin{equation}
    \frac{-(r-3\,\text{NM})}{min(-6\,\text{NM}/\text{sec}, \dot{r})} \le 60\,\text{sec}
\end{equation}
where $r$ is the slant range and $\dot{r}$ is the rate of change of slant range, with negative values occurring for decreasing $r$. 

In both cases, the active surveillance range is smallest when the two aircraft are effectively flying away from each other---here, the denominators would both default to the static values in the minimization functions (due to positive $\dot{s}$ and $\dot{r}$), forming a protection area around the aircraft. This is the worst-case scenario for the attacker; the smallest region in which they must be located relative to the aircraft in order to be under active surveillance. Through rearranging both equations, we can see that the vertical \textit{tau} region would be $s \le 4560\,\text{ft}$ and the horizontal would be $r \le 3.1\,\text{NM}$. If the target aircraft was instead flying towards the attacker, the denominators would be larger and so effectively increase the size of the protected area and reduce the constraints on the attacker---however, they will maximize their chance by being in the smallest area.

To have the best chance of successfully attacking CAS, the attacker must be within this protected region. As discussed above, when using Mode S the attacker has the ability to self-report altitude. This means that they can ensure they meet the first condition by injecting appropriate altitudes, or in the worst case simply matching the target's altitude.

Falling within the horizontal \textit{tau} region is more difficult, however. If we take the base case of $r \le 3.1\,\text{NM}$, or  $r \le 18836\,\text{ft}$, this could represent any attacker-aircraft position from the aircraft being 18836\,ft overhead to it being at low altitude almost 3\,NM horizontally away. 

If the attacker needed to inject a small burst of data, this would not matter. However, attacking CAS is different to attacking surveillance systems on their own. Since the system maintains state over time means that the attacker must be able to inject messages over a prolonged period---ideally as long as possible. Because of this, the attacker's chance of success is increased if $h$ is larger than $s$ for as long as possible. The most extreme case of this is where the attacker and target form a right angled triangle, which is at an altitude of 13306\,ft and range of 2.19\,NM. The other end of the spectrum is where the aircraft is at a minimum height where CAS is available, nominally 2350\,ft, which occurs at a range of 3.08\,NM. Using this we can establish a basic `vulnerable' altitude band in which the target and attacker are within the active surveillance range.

\section{Simulating Attacks on Collision Avoidance}
\label{sec:simulate}

Having identified a vulnerable range of altitudes, we now must test if this theory holds against CAS in practice. As such, we want to investigate whether attackers can inject false aircraft which cause RAs, and whether target aircraft in the range 2350-13306\,ft are particularly susceptible to attack. To do this, we use a simulator built on standardized CAS code from the ACAS X standard~\cite{RTCA2018}. In this section, we outline our simulator approach to experimentally test the vulnerability of ACAS X and the constraints on the attacker.

\subsection{Simulator Implementation}

\begin{figure}
    \centering
    \includegraphics[width=0.6\columnwidth]{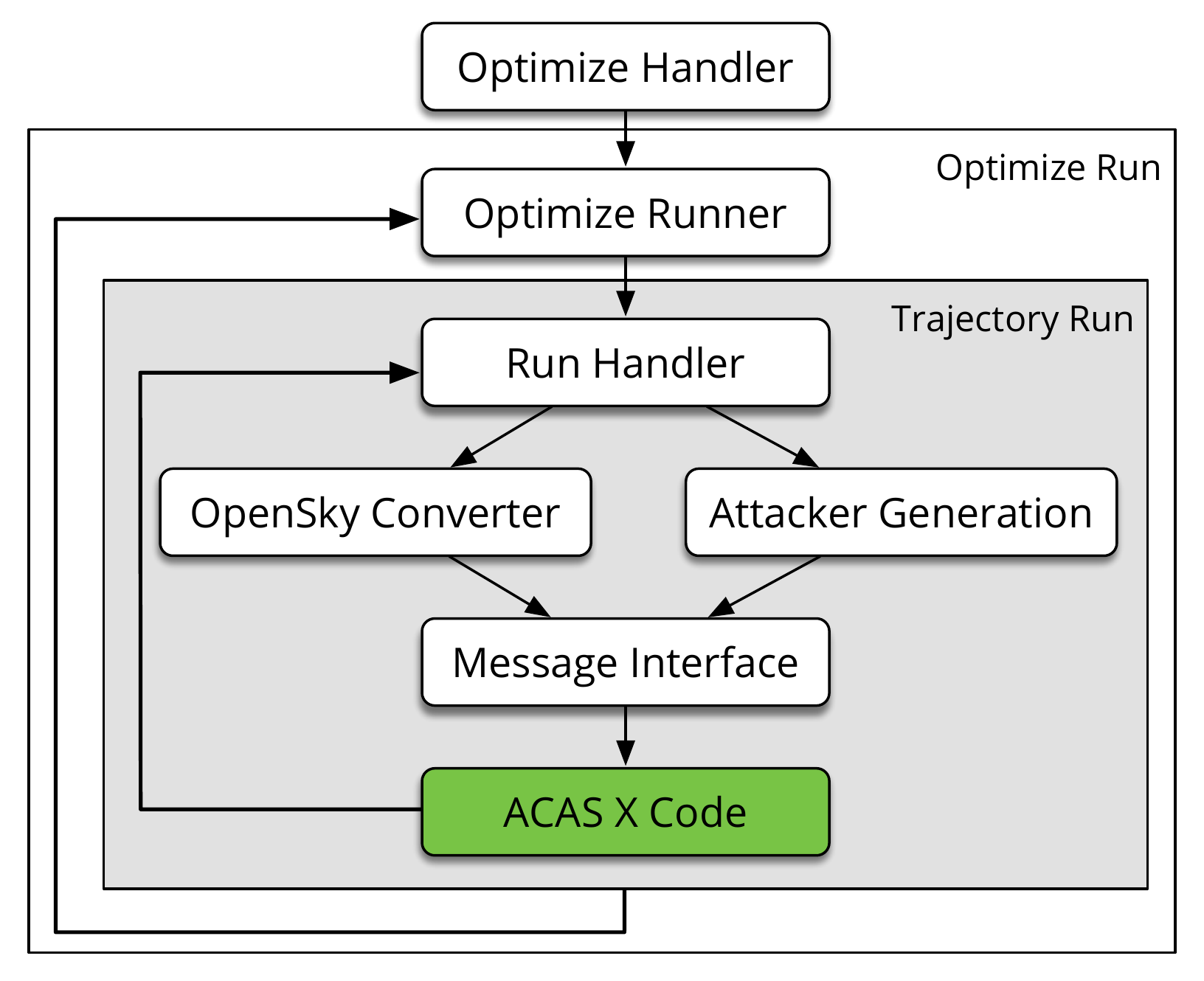}
    \caption{System diagram of our ACAS X simulation environment. Core ACAS X code is from DO-385 standard~\cite{RTCA2018}.}
    \label{fig:sysdiagram}
\end{figure}

One of the critical aspects of our work is representativeness of real collision avoidance systems. TCAS, although widely deployed, requires the implementation from standardized pseudocode. One of the aims of ACAS X was to reduce the implementation uncertainty this introduced, by using Julia rather than pseudocode, meaning that the specified implementation can be executed~\cite{Moss2015}. This can also be tested for correctness using a standardized a testing suite~\cite{Castle2012d}. We use this Julia code as our simulator core, which has been checked to ensure it passes all prescriptive (compulsory) tests provided in the standard. 

We wrapped this standardized code in our simulator harness, depicted in Fig.~\ref{fig:sysdiagram}. Within a single run, the \textit{run handler} converts an input trajectory for the ownship into messages accepted by ACAS X and combines it with intruder messages. We discuss how we generate both the ownship trajectories and input from intruders (namely attackers) below. The run handler does this conversion for each step, nominally one second, of a trajectory in the same way that a transponder would feed messages into ACAS X once per second.

The run handler controls the simulator during a single simulator run over a trajectory. We further wrap this in an \textit{optimization harness}, which allows us to do repeated simulation runs on a single trajectory, varying attacker parameters in order to find as optimal an attack as possible. This part of the simulator implements batch gradient ascent with random restart, thus repeatedly running according to optimization parameters. We discuss the details of our optimization approach below.

Testing against ACAS X in this way has an additional benefit. With the introduction of the cost table approach to decisions, it is much harder to `dry run' scenarios than with TCAS. Here, we can simply construct a trajectory and test the effects of attacks.

\subsection{Target Aircraft Trajectories}

In order to have a realistic target aircraft, we used flight trajectories extracted from the OpenSky Network over the course of 31 days between 15th Nov. and 15th Dec. 2019. We gathered the ADS-B data of flights falling within an approximately 80\,km diagonal bounding box centered on a given airport, for 6 airports: London Heathrow, Amsterdam Schipol, Frankfurt Am Main, New York John F. Kennedy and Washington Dulles. Due to the noisy nature of ADS-B, the data requires cleanup.

Concretely, we remove trajectories with more than 20\% of their barometric altitude reports missing and linearly interpolate any smaller gaps. Trajectories are further checked for discontinuities in altitude or excessively high rates of climb or descent and discarded if these features are present. Further details can be found in App.~\ref{app:pipeline}.

Once cleaned, we then treat this data as the input to the `ownship', i.e. the ACAS X simulator runs from the point of view of the aircraft for which we have the trajectory.

\subsection{Staging Attacks}

With a realistic ownship trajectory in place, we now detail how our attacker messages are constructed and how we attempt to find the most effective attacks for a given trajectory. 

\subsubsection{Attacker Behavior}
As described in our threat model, we implemented a static attacker relative to the aircraft who responds to target aircraft interrogations. We consider two attack positions, under the middle or end coordinate of the aircraft trajectory. This balances the complexity of the simulation while also providing an ideal, head-on position for the attacker. 

\subsubsection{Attacker Inputs}

As the simulator code itself treats incoming Mode S messages as if coming from the transponder, with range and bearing already estimated, we must carry out calculations of what the aircraft-derived measures would be based on the attacker and aircraft positions. This ensures that the simulation results are realistic---arbitrary bearings and slant ranges as inputs will certainly create positive results, but these will not map to the real world where the attacker cannot do this. 

With attacker position and aircraft altitude known, we can calculate the real slant distance and bearing based on an attacker response. Since Mode S allows the attacker to self-report altitude, this value can be picked arbitrarily. However, this will affect the calculated horizontal distance---as discussed in Sec.~\ref{sec:bounds}, a small altitude difference with a large slant distance leads to a similarly large horizontal distance. Throughout the attack, our attacker uses altitudes relative to the target aircraft, allowing them to `follow' the target. This guarantees that the injected aircraft will cross the path of the target aircraft and keeps it within the vertical protected volume. 

\subsubsection{Target Behaviour}
Since a successful attack may require the target aircraft to change its altitude, we must accommodate some response to that in order to get accurate results. We make some simplifying assumptions which capture the response without the finer details of pilot input. Specifically, if an RA requires a vertical rate change, we presume the target follows this immediately then returns to their original intention. For example, if the target is flying level then receives a climb RA, they follow the climb then continue to fly at their new level.

After an RA in the real world, ATC would contact the aircraft involved and issue instructions about how to proceed. In a simulated environment with two aircraft over a short period of time, we do not need these more complex instructions. 

\begin{figure}[t]
    \centering
    \subfloat[Single peak/plateau trajectory cost.]{
        \includegraphics[width=0.47\columnwidth]{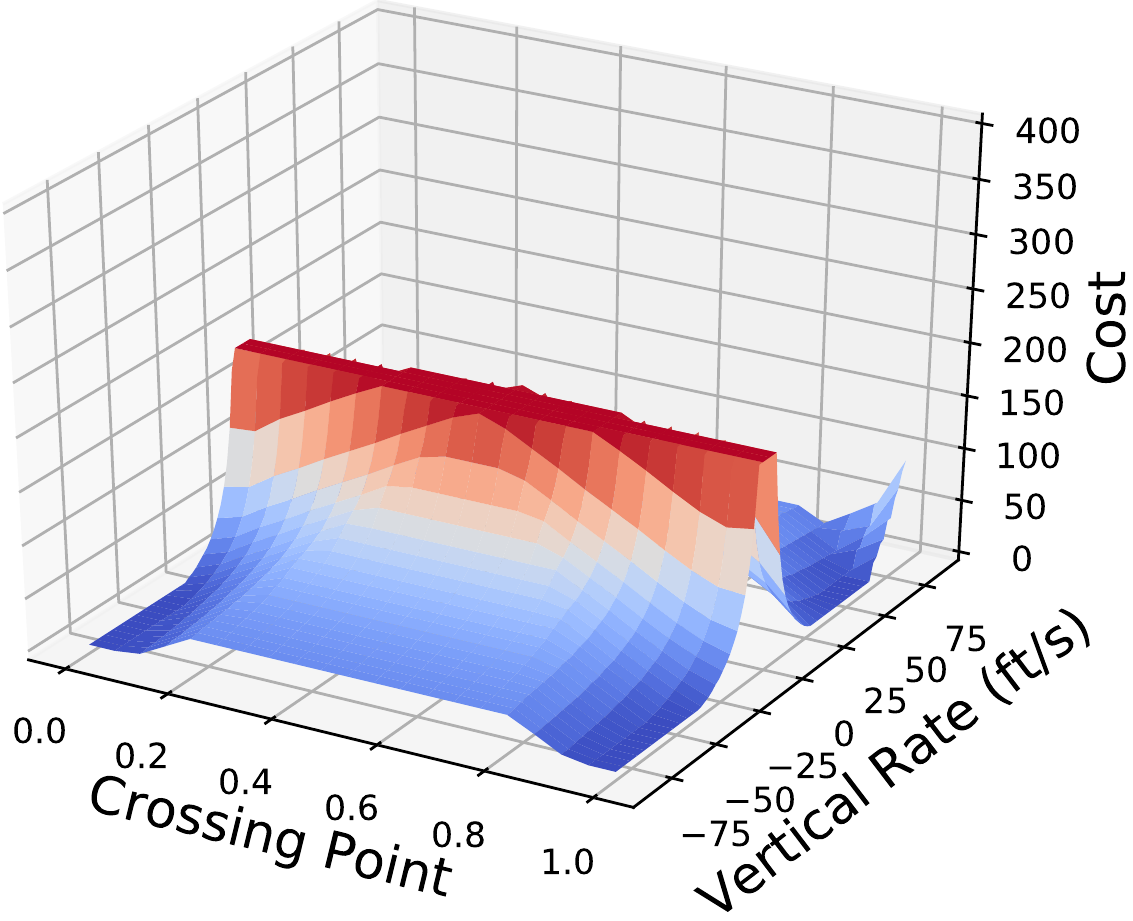}
        \label{fig:costmap-single}
    }\hspace{0.025cm}
    \subfloat[Multiple peak/plateau trajectory cost.]{
        \includegraphics[width=0.47\columnwidth]{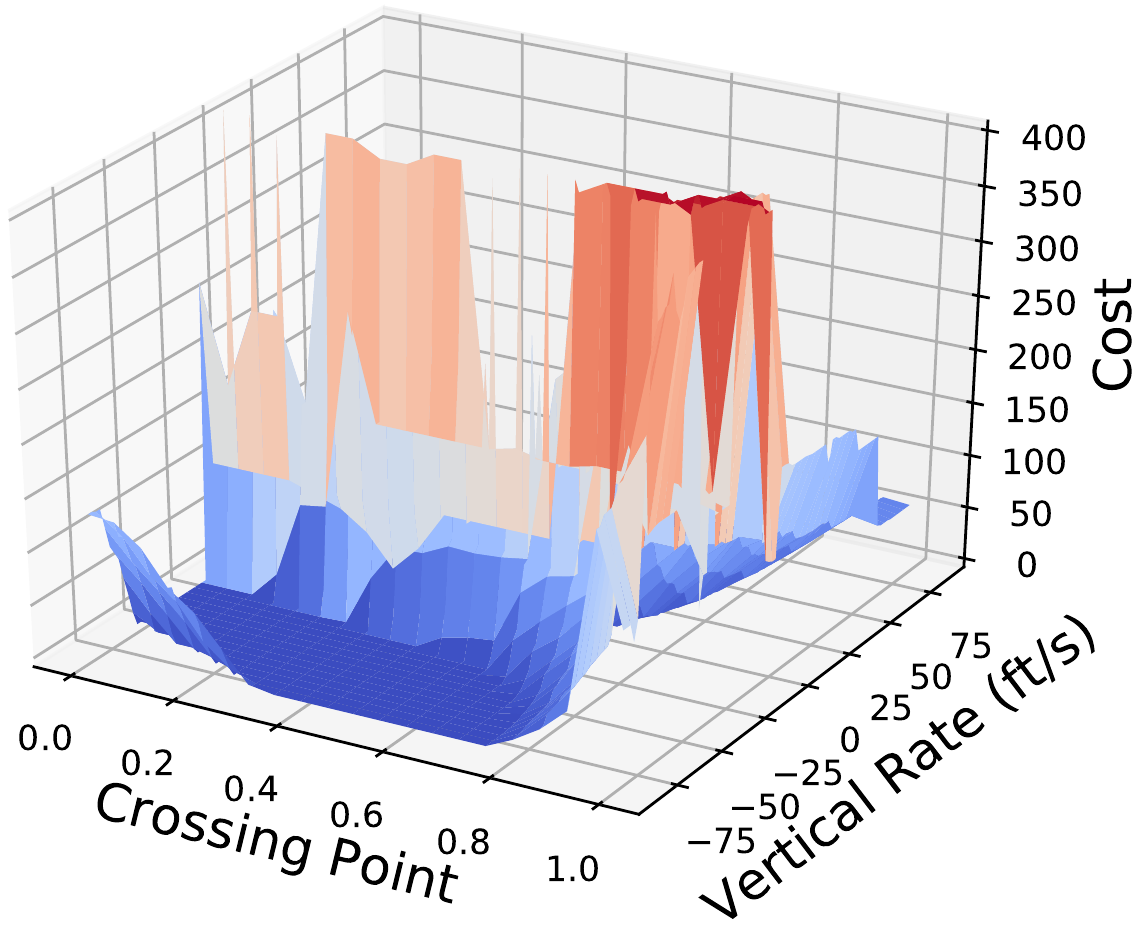}
        \label{fig:costmap-multiple}
    }
    \caption{Example cost maps across attacker optimization parameters for two trajectories, comparing how some trajectories have a clear best cost whilst others have multiple peaks.}
    \label{fig:costmap}
\end{figure}

\subsubsection{Optimisation of Attack}
While we have a broad idea of trajectories which might cause CAS alarms, the complexity of collision avoidance rules make it hard to know where the `best' attack lies. As we do not know the ideal attack positions, hand picking attack strategies will not give a good representation of how successful an attacker can be against a given trajectory. To find a good attack strategy for each trajectory, we used batch gradient ascent with random restart. We allow up to 20 iterations per trajectory---by inspection, runs hit local maxima multiple times in this length. We run each phase until the cost did not increase for some number of iterations, to allow traversal along cost function edges, or only decreases. At this point, we restart with a random strategy.

The injected aircraft follows a linear trajectory during runs, moving from a start altitude to end altitude at a given rate. Note that the attacker transmits towards the aircraft so the injected aircraft will be at the same bearing as the attacker to the target. We allow the optimization to vary three parameters:
\begin{itemize}
    \itemsep-1pt
    \item \textit{Crossing point}, the point in the run where the attacker crosses the target's altitude, i.e. is at the same altitude, between 0 and 1 where 1 is the final position of the aircraft in the trajectory.
    \item \textit{Rate}, or the vertical rate of change of altitude per simulation cycle, in feet per second, between -84 to 84. This corresponds to high rates of climb or descent for passenger aircraft based on data from the Eurocontrol Aircraft Performance Database~\cite{Eurocontrol2020}. This is referred to as vertical rate.
    \item \textit{Attacker position}, as a choice between two positions: underneath middle or end position of the target trajectory.
\end{itemize}
We start each optimization run with a random strategy and at each step calculate the cost of small changes in each of these three parameters. We use an exponentially decreasing learning strategy for each phase, resetting on a random restart.

\subsubsection{Measurement Criteria}

In order to optimize as above, we defined a simple cost function based on the severity of alerts in CAS. We use several metrics which help to quantify the length and impact of an attack. In particular, we are interested in how long:
\begin{enumerate}
    \itemsep-0.75pt
    \item the injected aircraft is proximate, $t_{PA}$,
    \item the target is in a Traffic Alert state, $t_{TA}$,
    \item the target is in a Resolution Advisory state, $t_{RA}$,
    \item the longest continuous RA is, $l_{RA}$,
    \item RA is active and requires a vertical rate change, $t_{VR}$.
\end{enumerate}

These criteria are in ascending order of importance. For example, an attack strategy which causes an RA alarm for one step is a nuisance but unlikely to have safety consequences beyond distraction. However, if the attacker can induce long RAs with predictable rate changes, they may create dangerous situations for the target aircraft and others nearby.

Since RAs and non-zero vertical rate changes are highest priority, our cost function for some run $x$ can be expressed as:
\begin{equation}
    C(x) = 1t_{PA} + 2t_{TA} + 5t_{RA} + 5l_{RA} + 10t_{VR}
\end{equation}

As discussed above, we cannot predict the best possible attack, so using such a cost function allows us to measure this. In many cases, the trajectories have clear peaks or plateaus in the cost function as shown in Fig.~\ref{fig:costmap-single}; however, some trajectories have multiple separate peaks as in Fig.~\ref{fig:costmap-multiple}. Random restart helps us to avoid getting stuck on local maxima.

\vspace{-5pt}
\section{Vulnerability Analysis}
\label{sec:results}

\begin{table}[t]
    \centering
    \caption{Altitude trends over the course of trajectories, grouped into descending, flying level or climbing by airport.}
    \footnotesize
    \begin{tabular}{lllllll}
        \toprule
                      & \multicolumn{2}{c}{Descending} & \multicolumn{2}{c}{Level} & \multicolumn{2}{c}{Climbing} \\ \cmidrule(l){2-3} \cmidrule(l){4-5} \cmidrule(l){6-7} 
                      & \#             & \%            & \#          & \%          & \#            & \%           \\ \midrule
        Amsterdam     & 502            & 50.2          & 405         & 40.5        & 73            & 7.3          \\
        Frankfurt     & 494            & 49.4          & 390         & 39.0        & 116           & 11.6         \\
        New York JFK  & 539            & 53.9          & 242         & 24.2        & 219           & 21.9         \\
        Heathrow      & 554            & 55.4          & 278         & 27.8        & 168           & 16.8         \\
        San Francisco & 495            & 49.5          & 251         & 25.1        & 254           & 25.4         \\
        Washington DC & 537            & 53.7          & 304         & 30.4        & 159           & 15.9         \\ \midrule
        Total         & 2763           & 46.1          & 2103        & 35.0        & 1134          & 18.9         \\ \bottomrule
        \end{tabular}
    \label{tab:trajtrend}
\end{table}

We now present our analysis of the simulation output, focussing on the real-world potential for an attacker to successfully interfere with airborne CAS. As discussed above, we extracted data from OpenSky for six airports with comprehensive coverage: London Heathrow (LHR), Amsterdam Schipol (AMS), Frankfurt am Main (FRA), New York John F. Kennedy (JFK), San Francisco International (SFO) and Washington Dulles (IAD).

Across all airports, the input trajectories had a median run length of 236 steps, with a standard deviation $\sigma$ of 153 steps. Of these, 2763 were descending, 2103 were flying level and 1134 were climbing. We summarize these figures by airport in Tab.~\ref{tab:trajtrend}. The trajectories came from a wide range of altitudes, with a median starting altitude of 15976\,ft ($\sigma$: 8911\,ft) and ending altitude of 12976\,ft ($\sigma$: 9871\,ft). The median altitude over the length of all trajectories was 14501\,ft  ($\sigma$: 9179\,ft).

\subsection{Attack Success}

\begin{figure}[t]
    \centering
    \includegraphics[width=0.9\columnwidth]{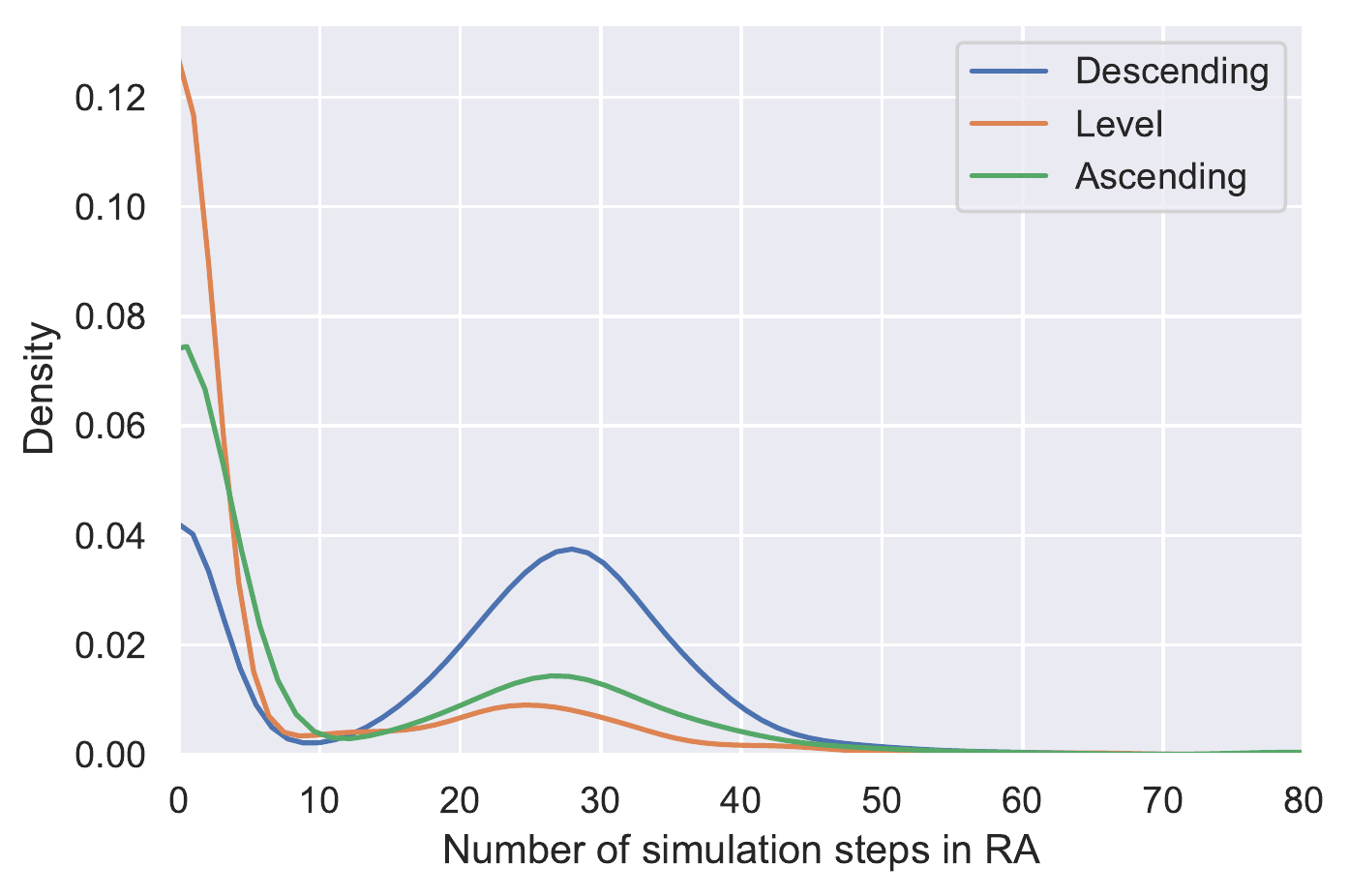}
    \caption{Probability density plot for the number of simulation steps spent in an RA, separated by aircraft descending, level and climbing. Descending aircraft more likely suffer RAs.}  
    \label{fig:RA-dir-pd}
\end{figure}

\begin{table}[]
    \centering
    \caption{Summary of the number of runs which, with their best scoring attacks, had instances of TAs, RAs and RAs with a non-zero target vertical rate and were partly contained in our modelling constraints. Percentages are of all trajectories.}
    \footnotesize
    \begin{tabular}{@{}lllllllll@{}}
    \toprule
     & \multicolumn{2}{c}{Has TAs} & \multicolumn{2}{c}{Has RAs} & \multicolumn{2}{c}{Vert. Rate} & \multicolumn{2}{l}{Partly Cont.} \\   \cmidrule(l){2-3} \cmidrule(l){4-5} \cmidrule(l){6-7} \cmidrule(l){8-9}
     & \# & \% & \# & \% & \# & \% & \# & \% \\ \midrule
    AMS & 614 & 61.4 & 568 & 56.8 & 561 & 56.1 & 618 & 61.8 \\
    FRA & 491 & 49.1 & 411 & 41.1 & 380 & 38.0 & 471 & 47.1 \\
    JFK & 515 & 51.5 & 387 & 38.7 & 218 & 21.8 & 489 & 48.9 \\
    LHR & 798 & 79.8 & 630 & 63.0 & 554 & 55.4 & 725 & 72.5 \\
    SFO & 617 & 61.7 & 457 & 45.7 & 372 & 37.2 & 659 & 65.9 \\
    IAD & 229 & 22.9 & 187 & 18.7 & 158 & 15.8 & 259 & 25.9 \\ \midrule
    Total & 3264 & 54.4 & 2640 & 44.0 & 2243 & 37.4 & 3221 & 53.7 \\ \bottomrule
    \label{tab:summary-stats}
\end{tabular}
\end{table}

At a high level, 3264 (54.4\%) of tested trajectories had one or more simulation steps with an active TA, 2640 (44.0\%) had one or more steps with an active RA and 2243 (37.4\%, or 85.0\% of the active RA steps) saw the CAS require a non-zero vertical rate. As we discuss below, aircraft not falling into this group also are impacted as they may have to level off, i.e. have a target vertical rate of 0. Thus, while an attacker is unlikely to be successful for every trajectory, when they can trigger an alert, there is a high likelihood of a kinetic effect.

By airport, Heathrow was the most vulnerable (see Tab.~\ref{tab:summary-stats}), with 63.0\% of runs having at least one RA and the vast majority of those RA runs resulting in a non-zero target vertical rate requirement (87.9\%). We investigate this further below. Amsterdam is the next most vulnerable in terms of RAs  (56.8\%), with Washington DC having the lowest percentage of runs result in an RA at 18.7\%. 

\subsubsection{Trajectory Altitude Trend}

While airports show some difference in attacker success, the altitude trend of trajectories, namely whether the aircraft is climbing, flying level or descending, has a clearer split. In Fig.~\ref{fig:RA-dir-pd}, we provide a probability density plot for the number of simulation steps spent in an RA, with a line for each altitude trend. First, it confirms that many flights have no, or very few, cycles in an RA. However, of the trajectories which do result in RAs, the vast majority are for descending aircraft. Looking to where most of the trajectories lie, in the 10-45 steps with an RA, a descending trajectory is over twice as vulnerable.

When considered in context, this is the scenario where an aircraft is flying most directly towards an attacker, which rapidly shortens the distance between the attacker and the target. Here, a false aircraft injected into the CAS will cause it to believe that it is on a collision course. This suggests that the attack is particularly successful during approach, which is a high-workload time for pilots and a busy part of airspace.

\subsubsection{Verification of Theoretically Vulnerable Altitudes}

\setlength\tabcolsep{4.5pt}
\begin{table}[t]
    \centering
    \caption{Target altitude trends for aircraft fully or partly (i.e. starts or ends) within theoretically vulnerable bounds.}
\footnotesize
\begin{tabular}{@{}llllllllll@{}}
    \toprule
           & \multicolumn{3}{c}{\begin{tabular}[c]{@{}c@{}}Partly \\ Contained\end{tabular}} & \multicolumn{3}{c}{\begin{tabular}[c]{@{}c@{}}Fully \\ Contained\end{tabular}} & \multicolumn{3}{c}{\begin{tabular}[c]{@{}c@{}}All\\ Trajectories\end{tabular}} \\ \cmidrule(l){2-4}\cmidrule(l){5-7}\cmidrule(l){8-10} 
            & RAs & Runs & \% & RAs & Runs & \% & RAs & Runs & \% \\ \midrule
            Desc. & 1793 & 2177 & 82.4 & 1350 & 1571 & 85.9 & 1872 & 2763 & 67.8 \\
            Level & 398 & 660 & 60.3 & 397 & 658 & 60.3 & 423 & 2103 & 20.1 \\
            Climb. & 271 & 384 & 70.6 & 192 & 227 & 84.6 & 345 & 1134 & 30.4 \\ \midrule
            Total & 2462 & 3221 & 76.4 & 1939 & 2456 & 78.9 & 2640 & 6000 & 44.0 \\ \bottomrule
            \end{tabular}
\label{tab:altdir-bounds}
\end{table}
\setlength\tabcolsep{6pt}

In Sec.~\ref{sec:bounds} we modelled that aircraft in the altitude range 2350-13306\,ft are the most exposed to injection attacks. Our results from the ACAS X simulator support this. Across successfully attacked trajectories, the median start altitude was 9775\,ft ($\sigma$: 4790\,ft) and end altitude was 6875\,ft ($\sigma$: 4382\,ft). The median of averages across each trajectory was 8488\,ft ($\sigma$: 4149\,ft), which along with the start and end altitudes, is comfortably inside the vulnerable range. We summarize the results, along with the relevant altitude trends, in Tab.~\ref{tab:altdir-bounds}, denoting both trajectories which start and end within the bounds (\textit{fully contained}) and those which start or end within the bounds (\textit{partly contained}). Note that partly contained is a superset of fully contained.

The results show aircraft are more vulnerable in this range. For trajectories both starting and ending within 2350-13306\,ft, 1939 of the 2456 (78.9\%) matching trajectories suffered RAs. For those, which either start or end within it, 2462 of 3221 (76.4\%)  trajectories suffered RAs. Both are almost twice as vulnerable compared to the full set of trajectories (44.0\%). 

Taking into account altitude trends, descending trajectories saw an 18 percentage point increase over all trajectories, while climbing and flying level increased almost three times over (54\% and 40\% respectively). Only 79 descending, 25 level and 74 climbing trajectories that do not at least partly fall in our boundaries have RAs, indicating that aircraft at altitudes outside of this band are less susceptible to attack. 

These insights help explain the airport split in Tab.~\ref{tab:summary-stats}, where Washington DC has the lowest success rate for RAs at 18.7\%. We can see that it also has the fewest partly contained trajectories---slightly over half the next lowest, JFK.

\begin{figure}[]
    \centering
        \includegraphics[width=0.9\columnwidth]{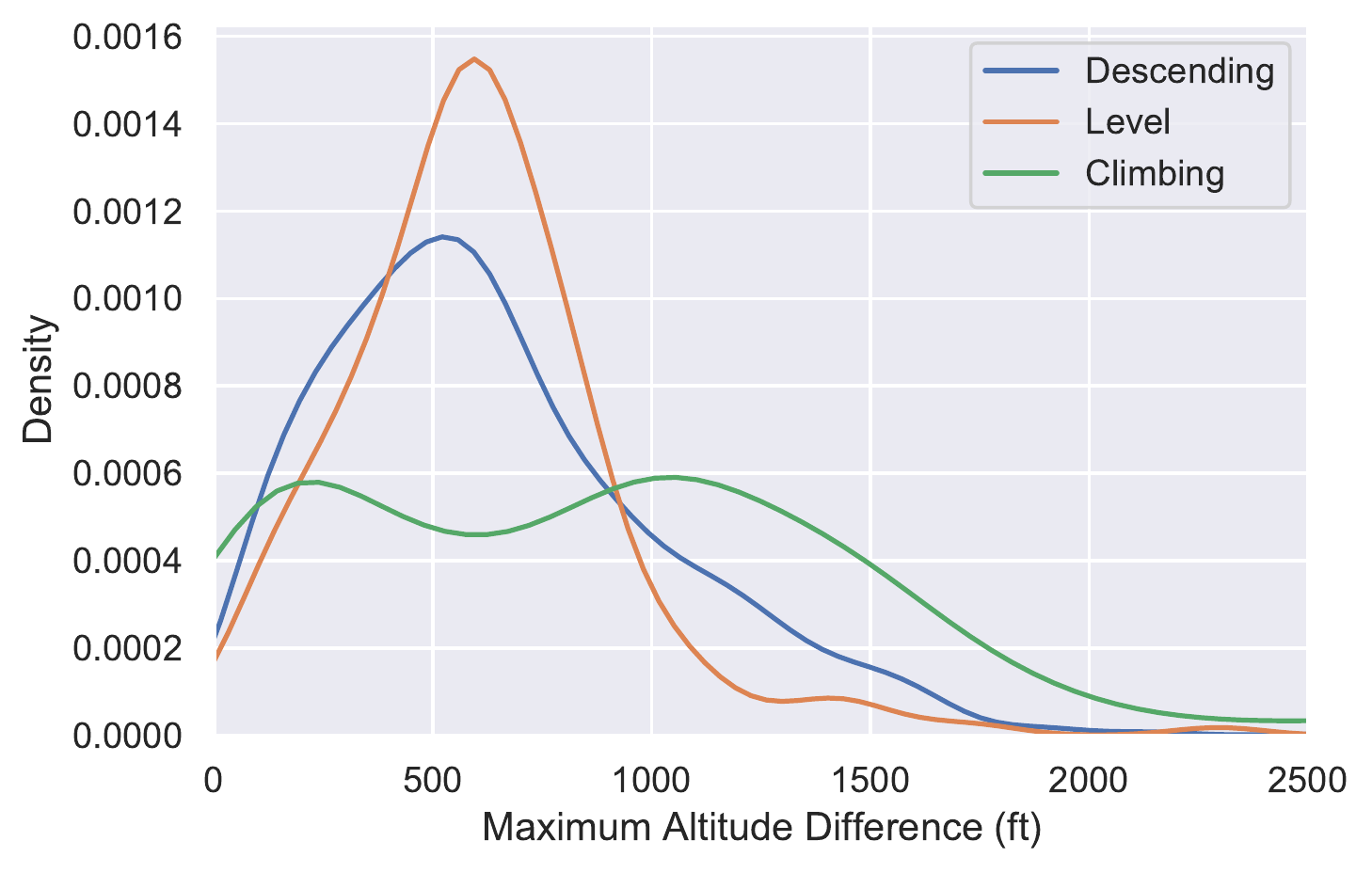}
    \caption{Probability density plot of the maximum difference between original and attacked trajectory for targeted aircraft. Descending and level aircraft have greater chance of causing deviations than climbing, but have lower average deviations.}
    \label{fig:altdir-maxdiff}\vspace{-5pt}
\end{figure}

\subsubsection{RA Length}
Looking more closely at the cases where attackers do generate RAs, we can measure the length of a typical continuous encounter that the attack might cause. Across all trajectories, the median `longest run' of an RA was 25 steps with a standard deviation of 14, where one step is one second. The ratio of `longest RA run length' to the total length of the trajectory for all trajectories had median value of 11.4\% with a standard deviation of 17.9\%. When compared to RA lengths in the real world, the absolute length matches quite closely---in analysis conducted by Eurocontrol, they found the average RA length to be 33 seconds~\cite{Drozdowski2009a}.

The trend of the trajectory appears to have some effect on its length. Climbing and level trajectories had the longest RA runs: climb had a median length of 26 steps ($\sigma$: 15), whereas descending a median length of 25 ($\sigma$: 14). Level trajectories had slightly shorter runs on average---a median length of 22 steps ($\sigma$: 16), possibly as a result of higher closing speeds whilst flying at a given altitude, or not following a traffic pattern around the airport thus reducing exposure.

\subsection{Impact of Attacker-Induced RAs}

\begin{figure}[]
    \centering
    \includegraphics[width=0.8\columnwidth]{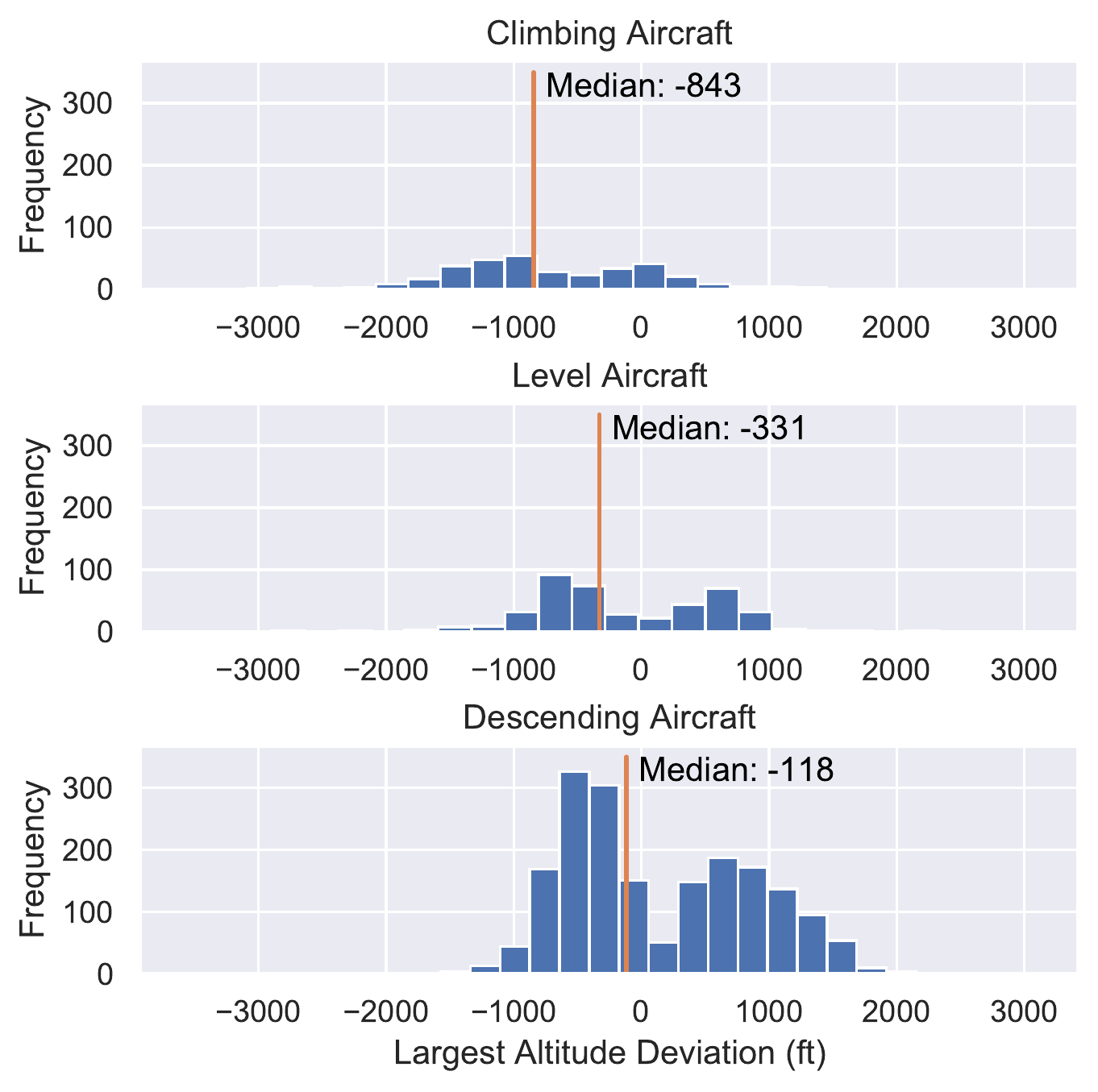}
    \caption{Histograms of the distributions of the largest altitude deviations across climbing, level and descending trajectories. Most attacks leave the target below its original trajectory.
    }
    \label{fig:greatest-diff-distro}
\end{figure}

Although triggering an RA could be seen as a success by itself, an attack is unlikely to have much effect if it does not cause any kinetic effect. We can measure this by comparing the trajectory of the best strategy against the original input trajectory, specifically looking at the altitude deviation. 

In Fig.~\ref{fig:altdir-maxdiff} we show the probability density of altitude displacement for trajectories in which the target aircraft has an RA for at least one simulation step. Descending and level target aircraft have similar distributions, with a median of 579\,ft and 584\,ft respectively. Climbing aircraft have a considerably higher median deviation at 901\,ft.

It does not appear that trajectories falling within the vulnerable bounds outlined previously have significantly different outcomes compared to all trajectories. For the fully contained trajectories, climbing trajectories saw a median maximum deviation of 804\,ft, with level at 584\,ft and descending at 559\,ft. Partly contained trajectories saw a slightly higher climbing median maximum deviation of 880\,ft, with level and descending at 584\,ft and 583\,ft respectively.

Under normal conditions (i.e. no CAS alert) this would cause a level bust, which is a situation where an aircraft deviates more than 300\,ft from its assigned altitude~\cite{SKYbrary2019}. Although these deviations may not be classed as a level bust since they are part of a collision avoidance maneuver, they would still incur the same potential harms. Specifically, this might cause the target aircraft to become too close to another nearby aircraft and cause further CAS alerts. It will also increase workload for air traffic controllers who may now have to adjust instructions for nearby aircraft.

Maximum altitude deviation alone does not describe the full impact. Fig.~\ref{fig:greatest-diff-distro} shows three histograms of the largest altitude deviations from the original trajectories, split by flight phase. The median greatest deviation for climbing aircraft was $-843$\,ft, level was $-331$\,ft and descending was $-118$\,ft. Trajectories falling outside of the vulnerable 2350-13306\,ft window differed; climbing trajectories had a median greatest difference of $-916$\,ft, level was $-409$\,ft and descending was 304\,ft. The larger difference in descending trajectories suggests that aircraft outside the window were more likely to have a level-off RA and end up above their original path.

\subsection{Attacker Strategy Analysis}

\begin{figure}[]
    \centering
        \includegraphics[width=0.9\columnwidth]{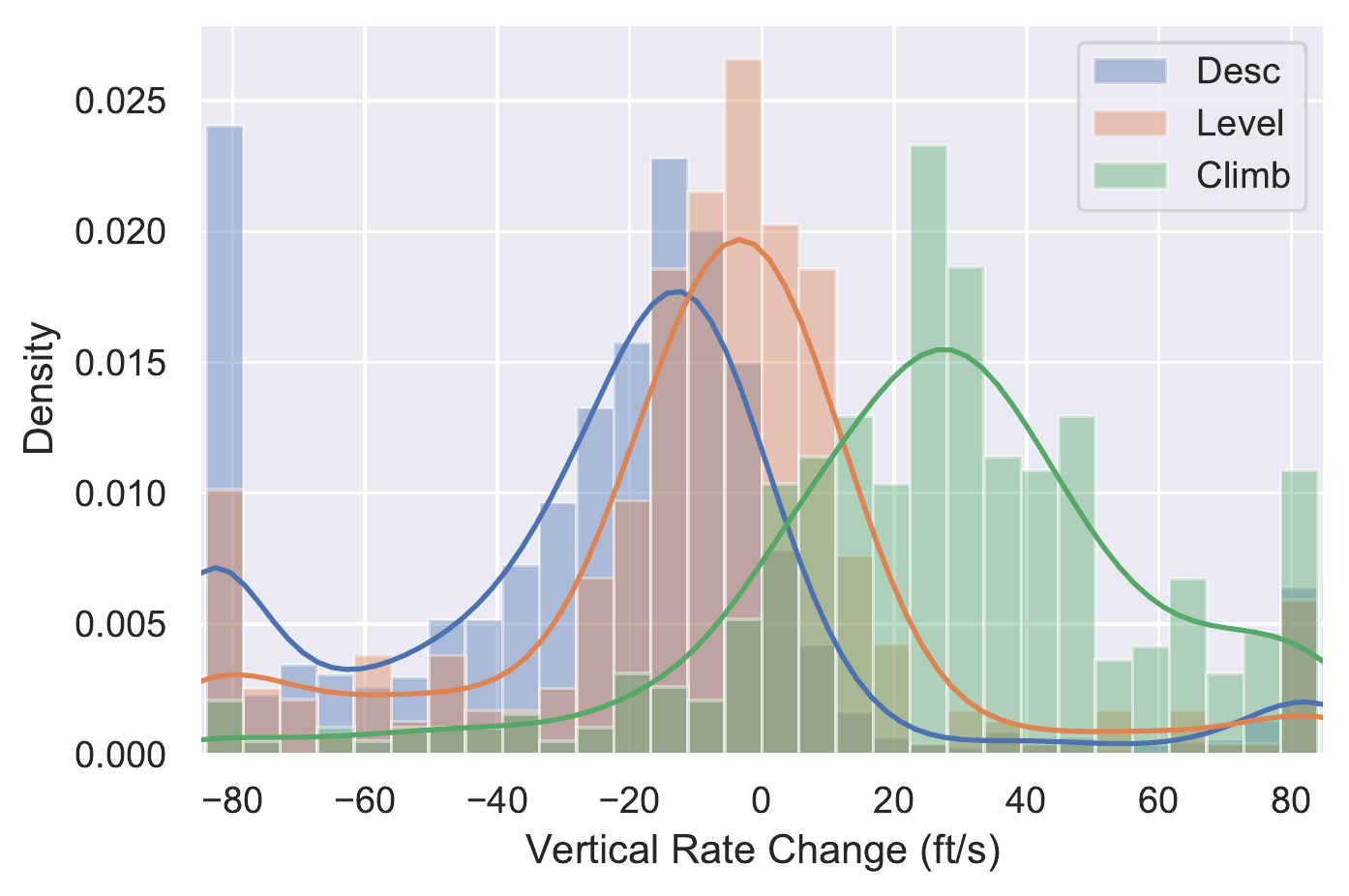}
    \caption{Density plot of the vertical rate change for the best strategy causing RAs. These are split into the altitude trends of climbing, flying level and descending. The median values show that the attacker must have a vertical rate in the same trend as the target to be successful. }
    \label{fig:rate-altdir-density}
\end{figure}

Finally, we look at the attacker-controller parameters used in the optimization process. This meta-analysis helps us to understand whether an attacker can pick an optimal position for a given trajectory to have the greatest chance of success. From a defender's point of view, this helps us to understand when an aircraft is most vulnerable to attack. We consider all trajectories with RAs as there was no significant parameter difference between those inside or outside of the bounds.

Over the full set of runs, trajectories suffering RAs had a median altitude change rate of $-12.9$\,ft/s ($\sigma$: 40.1), crossing point of 0.64 ($\sigma$: 0.29) with the attacker positions being split almost equally; 1424 (53.9\%) of successful runs were in the `middle' position, with the remaining 1216 (46.1\%) being at the end. In general, this suggests that an attacker injecting a descending aircraft which crosses the target aircraft's altitude around the half-way point of the target trajectory will give the best chance of success. However, as shown above, the success rate varies depends on trajectory characteristics. 

Crossing point and attacker position are strongly linked, with most RA-generating runs being clustered around the median for both end- and mid-positioned attackers. Successful mid-position attackers have a median position at 0.497 ($\sigma$: 0.221) and end-positioned are 0.985 ($\sigma$: 0.228), if we represent trajectories as unit length. Nominally, this corresponds to the crossover happening when the target is directly above the attacker. This is reasonable as it is also the point when the slant range between the attacker and the target is at a minimum.

\subsubsection{Vertical Rate Effect}

We previously explored the relationship between trajectory altitude trend and attacker success. Of the three parameters, vertical rate change is the most significantly affected and is presented in a density plot in Fig.~\ref{fig:rate-altdir-density}. Descending and level trajectories have the most RAs at rates just below 0, specifically medians of $-18.5$\,ft/s ($\sigma$: 37.7) and $-5.1$\,ft/s ($\sigma$: 33.4) respectively. Climbing trajectories have a much higher median rate change at 27.8\,ft/s ($\sigma$: 31.8). 

When the best attacker vertical rate for each trajectory is compared to the average altitude rate change for the input trajectory, the Spearman correlation coefficient is 0.482.\footnote{We use Spearman rather than Pearson as we do not know if the underlying altitudes are normally distributed and the barometric altitude reported is ordinal, quantized into at least 25\,ft steps~\cite{RTCA2011a}.} This positive correlation indicates that in many cases, the best attacking vertical rate has the same sign as the target vertical rate. In some senses, this is counterintuitive as one might expect RAs to be most common when where the injected aircraft flies `head on' to the target by having the opposite vertical rate.

\begin{figure}[]
    \centering
    \subfloat[Large $x$ when aircraft have the same vertical rate sign.]{
        \includegraphics[width=0.45\columnwidth]{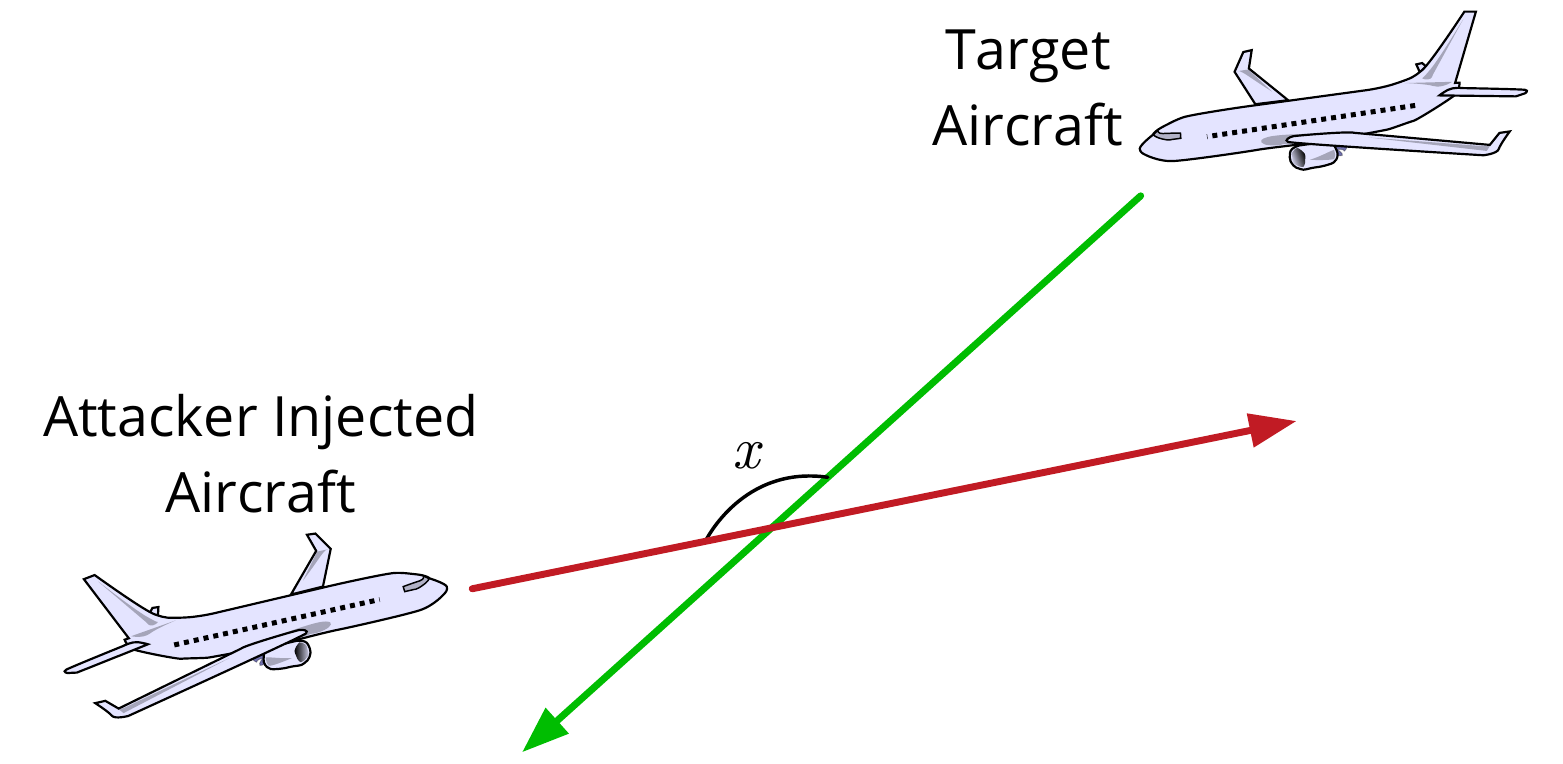}
        \label{fig:oppo-sign}
    } \hspace{0.01cm}
    \subfloat[Small $x$ when aircraft have opposite vertical rate sign.]{
        \includegraphics[width=0.45\columnwidth]{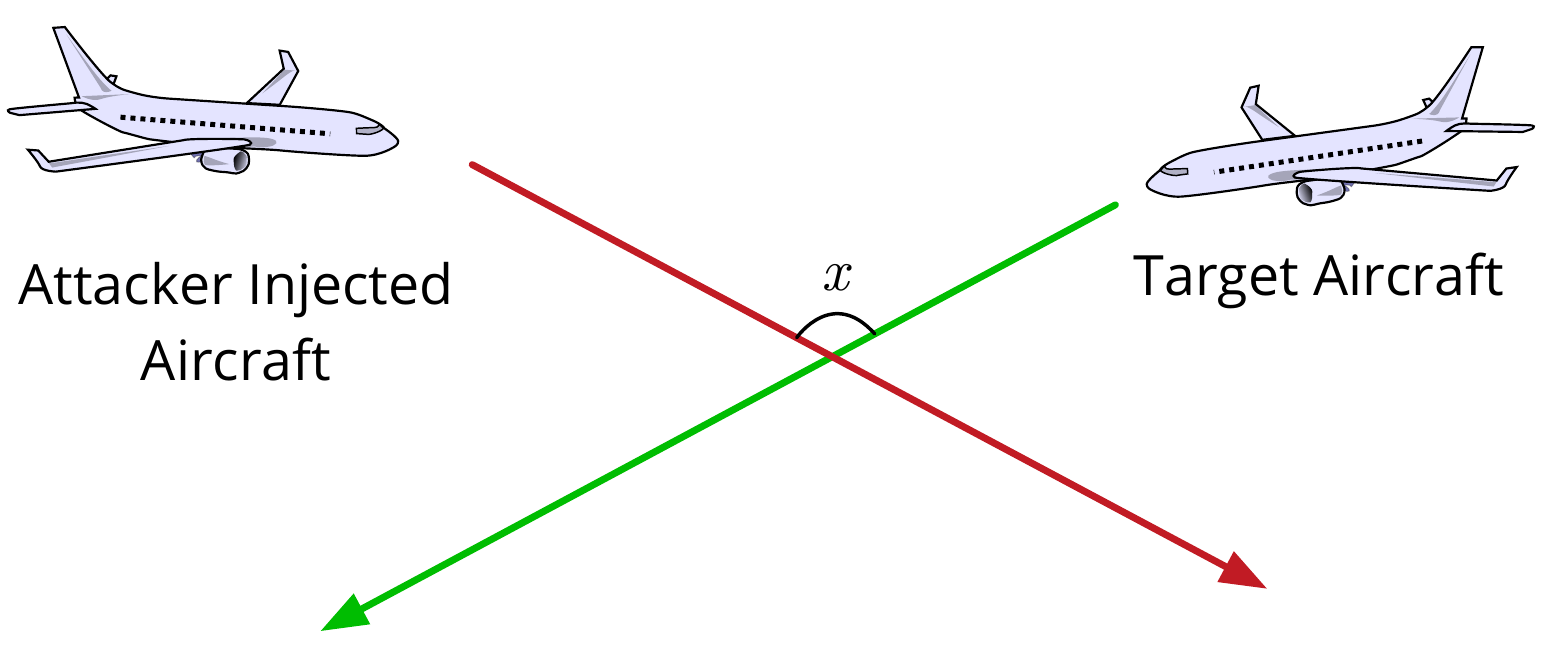}
        \label{fig:same-sign}
    }
    \caption{Effect on path angle between the target and attacker-injected aircraft when the attacker uses the a vertical rate with the same or opposite sign.}
    \label{fig:diag-crossing-angle}

\end{figure}
An explanation for this could be related to the amount of movement required by both aircraft to avoid a collision. If we consider the case where the injected and target aircraft are on paths which intersect, the target will need to adjust its path such that they no longer intersect. This involves reducing the intersection angle as quickly as possible, i.e. through following an RA. When the target and injected aircraft have opposite signs in their vertical rate as in Fig~\ref{fig:oppo-sign}, they have an obtuse angle between their intersecting paths. Deconflicting will involve increasing this angle over time until the aircraft are at least on parallel paths, which is a relatively small change. In contrast, when the signs are the same as in Fig~\ref{fig:same-sign}, the angle is acute and significantly larger changes are required to avoid the intersection. This causes greater deviation by the target to create separation, resulting in a long RA.

\section{Key Insights}
\label{sec:discuss}

\paragraph*{Attack Strategy}

We demonstrate that attacks on CAS are feasible and allow attackers to cause kinetic impact. There are significant variations in success chance depending on attacker position relative to the aircraft and aircraft behavior. Whereas random attacks have a success chance of 44\% in our simulations, it increases considerably to about 80\%, when the attacker focusses on aircraft flying at `vulnerable' altitudes.

Furthermore, we showed that the flight phase of the target aircraft---i.e. if it is climbing, descending or flying level---is correlated to attack success and impact type. We found that:

\begin{itemize}
    \itemsep-0.75pt
    \item Level aircraft would suffer unplanned climbs or descents,
    \item Climbing aircraft would often have to level off, resulting in a considerably lower than intended trajectory,
    \item Descending aircraft face a range of consequences, including having to level off, ending up above the planned altitude, or expediting descent and finishing below.
\end{itemize}

Of these scenarios, the descending aircraft is arguably the most concerning and formed the greatest proportion of our successful attacks. On such a path the aircraft is likely to be at vulnerable altitudes for longer than climbing or level aircraft. This increases both exposure to an attack and possible impact.

Finally, we found that the optimal attacker strategy is determined by the target trajectory. Our results suggest that attackers are most successful when injecting aircraft which cross the altitude of the target when the target is overhead. On top of this, the vertical rate change of the injected aircraft causes the greatest effect when it has the same sign to the target vertical rate. 

\paragraph*{Attack Effects}

A successful CAS attack has direct effects on the flight crew and the path of the aircraft. Across all successful attacks, the kinetic effect induced was enough to have wider consequences. Based purely on the magnitude of the deviations, an average attack would result in the aircraft being at least 500\,ft from its original trajectory, which is enough to cause a level bust. While level busts will only apply in some situations, ripple effects on the complex systems of busy airspaces are both likely and unpredictable. 

Although an attack focusses on a specific aircraft, its effects could be felt far beyond that. Considering that these RAs would arise from false aircraft not necessarily appearing on radar, this is likely to catch ATC by surprise. This may require a controller to have to immediately reorganize nearby airspace to ensure all aircraft are safe---especially in cases where RAs result in extreme climbing or descending.

Furthermore, the most vulnerable altitudes are commonly close to airports, which in turn means they are likely to be high-traffic sections of airspace. Aircraft are tightly controlled here and deviations could cause many aircraft to either receive adjusted ATC instructions or have CAS alerts of their own. An example of where this would be particularly concerning is in holding stacks, where aircraft fly patterns usually separated by 1000\,ft. An induced level bust because of an RA could cause many aircraft to have RAs in quick succession~\cite{Eurocontrol2019}.

The less immediate consequences also need to be considered carefully as they are of potentially even wider-ranging importance. As highlighted in previous work, pilots are effective at spotting unusual behavior in their systems but doing so draws their attention and diverts it away from other cockpit tasks~\cite{SmithNDSS2020}. If they feel the system is performing unusually, they are likely to reduce its sensitivity, switch it off or report it to ATC. CAS are important safety nets, having prevented many mid-air collisions globally since introduction \cite{schafer2019opensky}. Thus, effective denial of service attacks, which result in the system being turned off for the flight or undermine the trust in the system more generally are of potentially very high impact.
 
Similarly, if ATC notice many RAs occurring in an area, they can close that section of airspace or reroute aircraft---highly disruptive activities in and of themselves. Although both situations would stop the attacker having an effect, they also have considerable cost. Switching off CAS results in more work for ATC to keep the aircraft separated. Having to divert traffic could have knock-on effects including low-fuel incidents or diversions to other airports. 

\section{Countermeasures}
\label{sec:countermeasures}
As with many systems underpinned by avionic communications, adding cryptographic measures to the link would require system redesign, standardization and certification. This is both expensive and time consuming so would not address the issue in the short term \cite{Strohmeier2017}. 

An additional challenge exists for CAS, however. As with many safety systems, delayed or dropped messages have the potential to cause unsafe situations; more concretely, if two or more aircraft could not verify each others' identity and no fallback was allowed, collision avoidance would struggle to work as aircraft-to-aircraft communication would not be possible. Furthermore, both TCAS and ACAS X rely on passive observation of Mode C or S transmissions, meaning that the CAS must be able to verify any possible aircraft nearby too, whilst fulfilling its collision avoidance function.

Because of this, countermeasures to attacks on CAS must not impinge on their performance. Whilst a cryptographically protected CAS might be possible, in the shorter-term other defence mechanisms will provide immediate protection. We present two main forms: ground- and aircraft-based measures.

\subsection{Ground-based Measures}

\begin{figure}[]
    \centering
    \includegraphics[width=0.9\columnwidth]{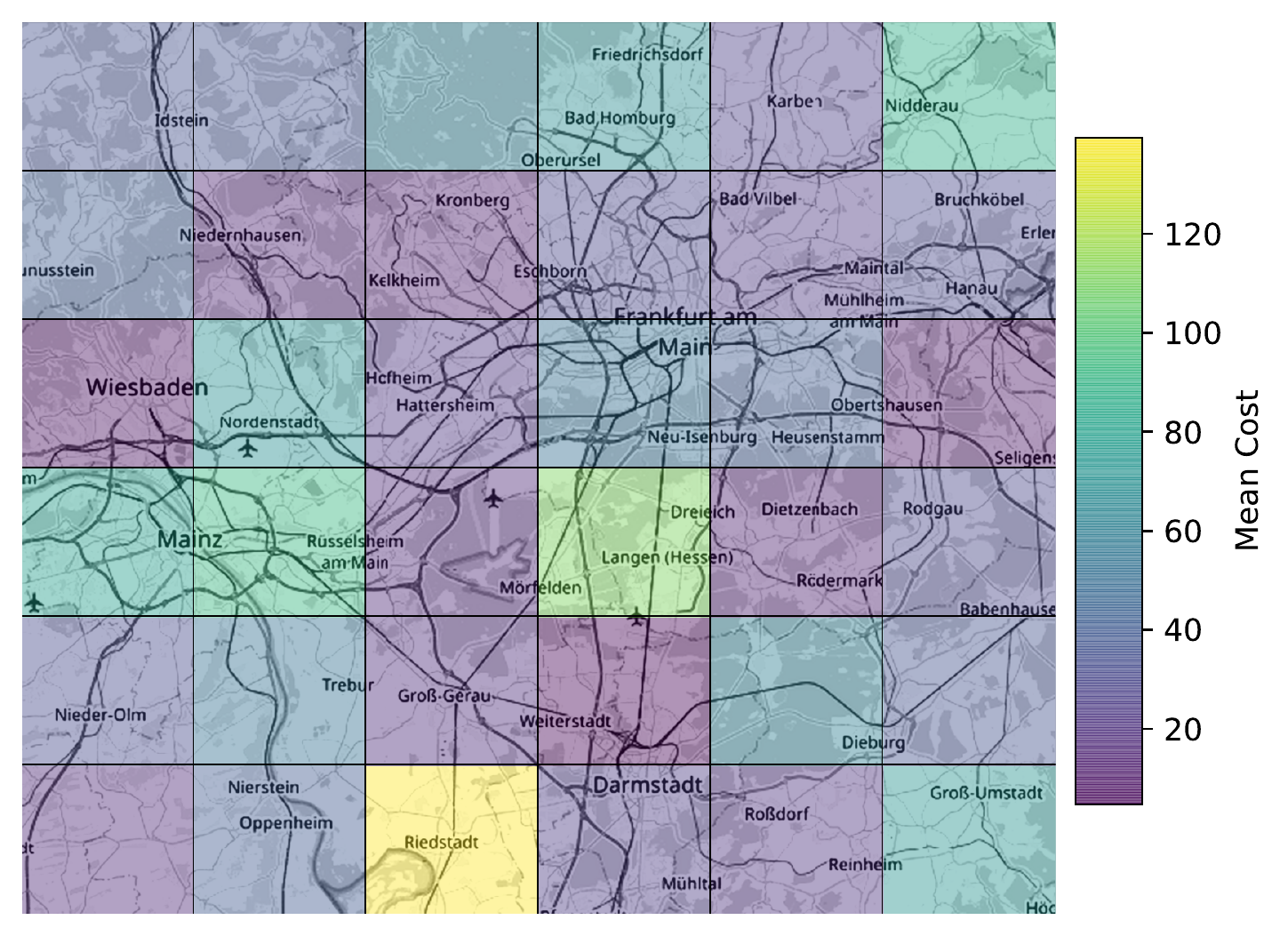}    
    \caption{Heatmap of cost function output for 100 trajectories at vulnerable altitudes around Frankfurt Airport.}
    \label{fig:heatmap}
\end{figure}

Carrying out the attack described in this paper requires a high transmission power over a sustained period of time---up to 500\,W at times to match the Mode S standard~\cite{RTCA2011}. Not only is such a transmission far higher power than would be allowed in the nearby spectrum, but the section of the spectrum itself is reserved for aviation~\cite{CNTIA2003,ECCECPTA2019,ICAO2018b}. Because of this, a monitoring system could identify the attacker transmissions and at least provide an alert that an attack could be happening. A more comprehensive system could provide some localization.

However, even if we presume the attacker is in the vicinity of an airport, the attack can be carried out over a wide area. Deploying sensors to comprehensively cover an area of around 3000\,$\text{km}^{2}$ would be extremely costly. Instead, we propose that our simulation approach can be used to model which areas are likely to be vulnerable to attack. By positioning the in-simulator attacker in a grid of points surrounding a given airport, we can produce costs for each grid position. Run across a range of trajectories, the simulator can produce an average cost per grid position over these flight paths.

To demonstrate this, we used a grid of 36 points distributed evenly around Frankfurt airport and within the bounds of our original data collection. We ran 100 trajectories classed as `fully contained' according to our theoretical bounding exercise and plotted the mean cost as a heatmap in Fig.~\ref{fig:heatmap}. With a limited run, we can identify areas with a higher average cost---specifically one towards the south stands out. Such analysis allows air traffic management to understand which approach or departure patterns might be vulnerable to attack. 

\subsection{Aircraft-based Measures}

Although monitoring and detection systems help to identify an attack in progress, this does mean that some aircraft will be under attack until the attacker can be located. To help mitigate this, some measures onboard the aircraft could be adopted. 

Some adaptations to the ongoing ACAS X implementation might be possible, including gathering more signal information and anomalous movement detection. At the signal level, CAS works separately from the transponder; the transponder handles the reception and decoding of a message before passing it over to CAS. As such, the CAS will have limited information about the underlying signal. One approach might be to provide more signal information either to CAS or a separate security device which monitors characteristics such as signal strength, timing, directionality and range.

Anomaly detection could be carried out with this richer information from the transponder. This might include monitoring intruders for unusual movement patterns, excessively high vertical rates or significant jumps in position. In our attacks, injected aircraft trajectories typically formed clusters of estimated positions above where the attacker was located. Fixed wing aircraft would not typically behave in this way. Importantly, we do not suggest dropping all anomalous data as we cannot be sure it is attacker-generated. Instead, this could be flagged to flight crew who would then be able to use the information when responding to any CAS alerts generated.

\section{Conclusion}
\label{sec:conclude}

In this paper we have identified that, subject to some altitude constraints, a suitably equipped attacker can carry out injection attacks on collision avoidance systems. By analyzing the workings of CAS systems, we highlighted that aircraft flying between the altitudes of 2350-13306\,ft are at the optimal combination of altitude and speed for a ground-based injection attack to be successful. We then tested this by harnessing standardized collision avoidance code from ACAS X, showing that across six airports in the US and Europe, an attacker had a 40-50\% chance of success for all aircraft in the area. However, when just considering the aircraft flying between 2350-13306\,ft, the attacker was considerably more successful---almost doubly so in some cases. 

The consequences of such an attack are significant. While causing mid-air collisions is unlikely, this attack causes direct disruption with the potential effects rippling out and effecting many aircraft nearby. We propose that to manage the risk of this attack, air traffic managers could use our simulation approach to map out high-risk areas and deploy monitoring systems there. Onboard the aircraft, the in-development status of ACAS X could allow for some additional security features to be considered, specifically around anomaly detection applied to intruder messages.

\bibliographystyle{plain}
\bibliography{main}

\appendix
\section{Trajectory Preparation Pipeline}
\label{app:pipeline}

Data as reported by aircraft and stored on the OpenSky Network can have irregularities such as altitude spikes or gaps with no reporting. This can be due to sensor quirks, processing delay or faulty reporting by the aircraft. For our simulator, we wanted clean input trajectories in order to monitor ACAS X behavior as accurately as possible. To produce these trajectories, we used a pipeline with the following steps:

\begin{enumerate}
   \itemsep-0.75pt
   \item Group flights by Mode S address and split into `flight trajectories', based on gaps of 60 seconds with no ADS-B reporting.
   \item Remove trajectories with more than 20\% of their barometric altitude reports missing. For remaining trajectories, linearly interpolate gaps between start and end values surrounding the gap. Gaps at the start or end of the trajectory are trimmed off.
   \item \label{item:trajectory-thresh}Trajectories are thresholded based on minimum and maximum altitudes, in order to remove unreliable low altitude reporting and high-altitude reporting. We included aircraft outside of our window of interest as defined in Sec.~\ref{sec:bounds} in order to test whether aircraft at other altitudes are vulnerable.
   \item \label{item:invalid-traj}Trajectories are further checked for discontinuities in altitude or excessively high rates of climb or descent, and discarded if these features are present.
\end{enumerate}

This pipeline produces trajectories with few disruptive artefacts due to ADS-B quirks. Specifically, Step~\ref{item:trajectory-thresh} removes extremely noisy data occurring below 3750\,ft where we found barometric altitude reporting to be significantly less accurate. We also remove cruise altitude aircraft, i.e. above 30000\,ft, as our bounding in Sec.~\ref{sec:bounds} identifies these as an unrealistic target. Similarly, those with unrealistic movements in altitude are removed in Step~\ref{item:invalid-traj}. We used a maximum climb of 5000 feet per minute and descent of 4500 as thresholds, based on surveying civilian aircraft performance data from the Eurocontrol Air Performance Database~\cite{Eurocontrol2020}.

\end{document}